\documentclass[prc,aps,nofootinbib,onecolumn]{revtex4}

\usepackage[english]{babel}
\usepackage[utf8]{inputenc}
\usepackage{color}
\usepackage{listings}
\usepackage{url}
\usepackage{amsmath,amsthm, amssymb, latexsym}
\usepackage{verbatim}
\usepackage{moreverb}
\usepackage[colorlinks=true]{hyperref}
\usepackage{graphicx}
\usepackage{xspace}
\usepackage{float}
\usepackage{bbm}
\usepackage{ifthen}
\usepackage{xifthen}
\usepackage{delimset}

\newcommand{\mathopl}{\mathcal}

\newcommand{\fvec}[1]         {\boldsymbol{#1}}

\newcommand{\diff}            {\operatorname{d}\!}

\newcommand{\Mvec}            {\mathbf{M}}

\newcommand{\Efunc} [1][]{\mathcal{E}\ifthenelse{\isempty{#1}}{}{[{#1}]}}

\newcommand{\vecRep}     [1]  {\mathrm{#1}} 
\newcommand{\opRep}      [2][]{\mathopl{#2}\ifthenelse{\isempty{#1}}{}{[#1]}}
\newcommand{\kerRep}     [2][]{\mathopl{#2}\ifthenelse{\isempty{#1}}{}{[#1]}}
\newcommand{\col}        [1]  {\tilde{\mathopl{#1}}}

\newcommand{\nop}        [1][]{\opRep[#1]{N}}
\newcommand{\sqrtop}     [2][]{\ifx#1p\left[{#2}\right]\else{#2{}}\fi^{\frac{1}{2}}}
\newcommand{\invsqrtop}  [2][]{\ifx#1p\left[{#2}\right]\else{#2{}}\fi^{-\frac{1}{2}}}
\newcommand{\nker}       [1][]{\kerRep[#1]{N}}
\newcommand{\hker}       [1][]{\kerRep[#1]{H}}

\newcommand{\hop}        [1][]{\opRep[#1]{H}}
\newcommand{\kop}        [1][]{\opRep[#1]{K}}
\newcommand{\overlap}         {\mathcal{N}}

\newcommand{\bra}        [1]  {\ensuremath{\left\langle{#1}\right|}\xspace}
\newcommand{\ket}        [1]  {\ensuremath{\left|{#1}\right\rangle}\xspace}
\newcommand{\braket}     [3][]{\ensuremath{\ifthenelse{\isempty{#1}}{\left\langle{#2}\middle|{#3}\right\rangle}{\delimtriple<\vert\vert>!{#2}{#1}{#3}}}\xspace }

\newcommand{\qavg}       [1]  {\ensuremath{\langle{#1}\rangle}}
\newcommand{\op}         [1]  {\hat{#1}}
\newcommand{\opId}            {\op{\mathbbm{1}}}

\newcommand{\HFB}             {\phi}
\newcommand{\bHFB}       [1]  {\bra{\HFB({#1})}}
\newcommand{\kHFB}       [1]  {\ket{\HFB({#1})}}
\newcommand{\bkHFB}      [3][]{\braket[{#1}]{\HFB({#2})}{\HFB({#3})}}

\newcommand{\kqpHFB}     [2]  {\ket{\HFB_{#2}({#1})}}
\newcommand{\bmHFB}      [1]  {\bra{\tilde{\HFB}({#1})}}
\newcommand{\kmHFB}      [1]  {\ket{\tilde{\HFB}({#1})}}

\newcommand{\TDGCM}           {\Psi}
\newcommand{\bTDGCM}     [1]  {\bra{\TDGCM(#1)}}
\newcommand{\kTDGCM}     [1]  {\ket{\TDGCM(#1)}}
\newcommand{\bkTDGCM}    [3][]{\braket[{#1}]{\TDGCM({#2})}{\TDGCM({#3})}}
\newcommand{\Qdim}            {m}
\newcommand{\SOPO}       [3]  {\left\{{#2} , {#3}\right\}^{(#1)}}

\newcommand{\ladder} [4]{\def\tempid{#1}\def\tempac{#2}\ifx\tempid\empty {#3}_{#4}\ifx#2c^{\dagger}\fi \else {#3}^{(#1)\ifx#2c{\dagger}\fi}_{#4}
                          \fi}
\newcommand{\idQPw}      {\eta}
\newcommand{\aQPw} [2][] {\ladder{#1}{a}{\hat{\idQPw}}{#2}}
\newcommand{\cQPw} [2][] {\ladder{#1}{c}{\hat{\idQPw}}{#2}}

\newcommand{\idPw}      {a}
\newcommand{\aPw} [2][] {\ladder{#1}{a}{\hat{\idPw}}{#2}}
\newcommand{\cPw} [2][] {\ladder{#1}{c}{\hat{\idPw}}{#2}}

\begin{document}

\title{The time-dependent generator coordinate method in nuclear physics}

\author{Marc Verrière} \email{verriere1@llnl.gov}
\affiliation{Los Alamos National Laboratory, Los Alamos, NM 87545, USA \\
  Nuclear and Chemical Sciences Division, Lawrence Livermore National Laboratory, Livermore, California 94551, USA}

\author{David Regnier$^*$} \email{david.regnier@cea.fr}
\affiliation{CEA, DAM, DIF, 91297 Arpajon, France \\
  Université Paris-Saclay, CEA, Laboratoire Matière en Conditions
  Extrêmes, Bruyères-le-Châtel, FRANCE \\
}

\begin{abstract}  

The emergence of collective behaviors and the existence of large 
amplitude motions are both central features in the fields of nuclear
structure and reactions. From a theoretical point of view, describing such
phenomena requires increasing the complexity of the
many-body wavefunction of the system to account for long-range correlations. 
One of the challenges, when going in
this direction, is to keep the approach tractable within our current
computational resources while gaining a maximum of predictive power for the
phenomenon under study.
In the Generator Coordinate Method (GCM), the many-body wave function is a linear superposition of
(generally non-orthogonal) many-body states (the generator states) labeled by a few
collective coordinates. 
Such a method has been widely used in structure studies to restore the
symmetries broken by single-reference approaches. In the domain of
reactions, its time-dependent version (TDGCM) has been developed and
applied to predict the dynamics of heavy-ion collisions or
fission where the collective fluctuations play an essential role.
In this review, we present the recent developments and applications of the
TDGCM in nuclear reactions. We recall
the formal derivations of the TDGCM and its most common approximate
treatment, the Gaussian Overlap Approximation.
We also emphasize the Schr\"odinger Collective-Intrinsic Model (SCIM) variant focused
on the inclusion of quasiparticle excitations into the description. 
Finally, we highlight several exploratory studies related to a TDGCM built on 
time-dependent generator states.
\end{abstract}

\maketitle

\tableofcontents

\section{Introduction}
\label{sec:intro}
Since the early days of nuclear physics, the variety of shapes that atomic
nuclei can take is a core notion of our interpretation of nuclear processes.
The fission reaction provides a typical example since it was quickly interpreted
as the elongation of a charged liquid drop of nuclear matter, leading to a
scission point~\cite{bohrMechanism1939}. Descriptions in terms of vibrations
and rotations of the nuclear shape also lead to quantitative reproductions
of the low energy spectra~\cite{bohrNuclear1998} of atomic nuclei.
These successes of the theory suggest that the shape of the nuclear density is
somehow a relevant degree of freedom (DoF) to describe several phenomena. In
addition to the classical picture of the time evolution of a well defined
nuclear shape, taking into account its associated quantum fluctuation is of
particular importance. For instance, these fluctuations directly drive the
width of the probability distribution of particles transferred during low
energy heavy-ion collisions, as well as the modal characteristics of the
fragment distribution produced by fission. The incorporation of these fluctuations
into a quantum description leads to a many-body wave function describing the system
that is a mixture of states with different shapes.
With this intuition, one may attempt a direct description of nuclei in terms of
shape DoFs. However, transforming the $3A$ positions and $A$ spins of the
nucleons into a new system of coordinates involving a set of deformations
parameters is both cumbersome and problem-dependent~\cite{scheidTheory1968,
finkSpurious1972}.
Another possibility consists in keeping the nucleons coordinates and build an
\textit{ad-hoc} quantum mixture of many-body states with different relevant
shapes. This is precisely the starting point of the Generator Coordinate Method
(GCM).

The GCM method was first developed in the seminal papers of Hill and Wheeler in
the context of nuclear fission in 1953~\cite{hillNuclear1953}, and later on
generalized in Ref.~\cite{griffinCollective1957}. The global philosophy is
(i) to generate a set of many-body states parametrized by a set of shape
variables (the generator states),
(ii) to derive an equation of motion for the many-body wave function of the
system in the restricted Hilbert space spanned by the generator states. 
The first applications of this method focused in introducing
shape degrees of freedom, such
as the multipole moments of the one-body density. It turns out to be
very versatile and has been applied since with different families of generator
states. The static GCM has demonstrated over the years its ability to describe
the low excitation spectrum of nuclei~\cite{benderGoing2008,
egidoStateoftheart2016}. For this kind of application, the generator states are,
in general, parametrized by some gauge variables associated with the breaking
and restoration of symmetry groups (Euler angles for rotational symmetry, gauge
angle for the particle-number symmetry)~. Similar approaches based on
generator states labeled by a few multipole moments of the one-body density
also provided predictions of the giant monopole, dipole, and quadrupole
resonances~\cite{krewaldSelfconsistent1976,stoitsovGenerator1994,
caurierMicroscopic1973,flocardGenerator1976,abgrallMonopole1975,
vretenarIsoscalar1997}.

Studies based on the time-dependent flavor of the GCM are less abundant in the
literature than the ones using its stationary counterpart. Therefore, the
goal of this review is to recall the formal developments related to the
Time-Dependent Generator Coordinate Method (TDGCM) and highlight their current
applications in the field of nuclear physics. In Sec.~\ref{sec:gcm}, we present
some general aspects of the time-dependent generator coordinate method in its
standard and full-fledged implementation.  In Sec.~\ref{sec:goa}, we focus on
the Gaussian overlap approximation framework that is commonly used in most of
the state of the art applications of the TDGCM. In particular, we discuss the
fact that such an approach has difficulties accounting for the diabatic aspects
of nuclear collective motions.  We then devote the two last sections to two
possible extensions of the TDGCM that aim to overcome this issue. The
Sec.~\ref{sec:scim} highlights the Schr\"odinger Collective Intrinsic Model
(SCIM), a framework based on the symmetric moment approximation of the TDGCM.
Finally, Sec.~\ref{sec:mctdhf} reports alternative methods involving a
TDGCM-like ansatz built on time-dependent generator states.
 
\section{General formalism of the TDGCM}
\label{sec:gcm}
\subsection{Generator states}
\label{subsec:gcm:coll}

Predicting the structure and dynamics of medium to heavy nuclei starting from
the nucleons degrees of freedom is a challenging task. The difficulty arises
from a large number of correlations present in the many-body wave function of
nuclear systems. A feature that helps us tackle this problem is the existence
of two nearly separable time scales in nuclear processes. On the one hand, we
have the typical time for the motion of individual nucleons inside the nucleus,
which is roughly 10$^{-22}$s. On the other hand, the time scales associated
with the system's collective deformations are roughly ten times bigger than the
former ($1$ zs = 10$^{-21}$s). Such separation in time scale motivated attempts
to describe the dynamics in terms of shape coordinates only. As mentioned in
the introduction, one possibility is to transform the 3$A$ positions of the
nucleons into a set of collective coordinates plus some residual intrinsic
DoFs. Such an approach could then be combined with an adiabatic approximation
similar to the Born-Oppenheimer approximation in electronic systems to reduce
the dynamics to the collective DoFs only. The GCM proceeds with an alternative
approach that introduces collective deformations DoFs without relying on a
transformation of the set of nucleons DoFs.

The first step of the method consists in building a family of many-body states
$\{\kHFB{\fvec{q}}\}$ parametrized by a vector of labels $\fvec{q}=q_0\cdots
q_{\Qdim-1}$. We can summarize the essence of such a construction in the
following few points:
\begin{itemize}
  \item The labels $q_i$ are referred to as the generator coordinates or
    collective coordinates. They are continuous real numbers that can, for
    instance, characterize the shape of the nuclear density. The vector
    $\fvec{q}$ takes arbitrary values in a $\Qdim$-dimensional subspace
    $E \subset \Re^{\Qdim}$.
 \item The states $\{ \kHFB{\fvec{q}} \}$ are the generator states. They are
   many-body states associated with the system of $A$ nucleons under study. In
    the standard TDGCM framework, these states are time-independent.
 \item The function $\fvec{q} \rightarrow \kHFB{\fvec{q}}$ should be
   continuous. In other words, for any sequence of collective coordinates
    $\{\fvec{q}_k\}$ that converges to $\fvec{q}$, the corresponding sequence
    $\kHFB{\fvec{q}_k}$ must converge to $\kHFB{\fvec{q}}$. This property is
    required for a sound mathematical construction of the GCM framework as
    detailed in Ref.~\cite{deKinematics1978}.
\end{itemize}
The choice of a family of generator states fulfilling these properties is then
arbitrary, which gives great versatility to the GCM method \footnote{For some
applications, it may be convenient to add one or several discrete generator
coordinates. We will then note the generator states as
$\kqpHFB{\fvec{q}}{\fvec{k}}$ where $\fvec{k}$ is a vector of discrete labels.
A typical example of a discrete label could be the $K$ quantum number
associated with the projection of the total spin onto a symmetry axis of the
nucleus. Another example is provided in Sec.~\ref{sec:scim}.}. The generator
states should span a sub-Hilbert space that contains each stage of the exact
dynamics to describe a physical process optimally. Therefore, building a
pertinent family of generator states requires a good \textit{a priori}
knowledge of the dynamics of the system.

A standard procedure to handle nuclear deformations consists in the definition
of the generator states as the solutions of a constrained
Hartree-Fock-Bogoliubov equation. In this approach, each collective coordinate
is typically associated with a multipole moment observable (i.e., the
quadrupole moment of the one-body density). The generator state
$\kHFB{\fvec{q}}$ is then obtained by minimizing the Routhian
\begin{equation}
 R[\phi(\fvec{q})] = E_{\text{HFB}}[\phi(\fvec{q})]
 - \sum_i \lambda_i
  \left(\bkHFB[\hat{Q}_i]{\fvec{q}}{\fvec{q}} - q_i \right)^2,
\end{equation}
where the $\hat{Q}_i$ refer to the chosen multipole operators and $\lambda_i$
are their associated Lagrange multipliers. This method presents the benefit of
controlling the principal components of the shape of the states through a small
set of DoFs. The other DoFs are determined automatically from the HFB
variational principle. It is often qualified as an adiabatic method because the
generator states will minimize their HFB energy under a small number of
constraints. One drawback of this method is that it does not necessarily ensure
the continuity of the function $\fvec{q} \rightarrow \kHFB{\fvec{q}}$. This
could severely affect some applications as mentioned in
Sec.~\ref{sec:tdgcm_appli} and \ref{sec:goa_limitations}.

In the context of nuclear structure, the now-standard strategy of symmetry
breaking and restoration provides a different yet natural way of building
generator states. In this context, we typically define the generator states as
the result of applying a parametrized group of symmetry operators on a
reference (and symmetry breaking) HFB state $\ket{\phi}$. Typically, for the
particle-number symmetry, the relevant collective coordinate is the gauge angle
$\theta$~\cite{ringNuclear2004} and the generator states $\kHFB{\theta}$ read
\begin{equation}
 \kHFB{\theta}  = \operatorname{exp}
  \left({i \theta(\hat{A}- A)}\right) \ket{\phi}.
\end{equation}
Note that the two strategies mentioned above to create the generator states are
often mixed when dealing with several collective
coordinates~\cite{egidoStateoftheart2016}.

\subsection{Griffin-Hill-Wheeler ansatz}
\label{subsec:gcm:GHWansatz}

Once the family of generator states is chosen, the Griffin-Hill-Wheeler (GHW)
ansatz assumes that the many-body state of the system reads at any time
\begin{equation}\label{eq:gcm:GHW}
  \kTDGCM{t} =
    \int_{\fvec{q}\in E} \diff{\fvec{q}}
      \kHFB{\fvec{q}}
      f(\fvec{q}, t).
\end{equation}
The function $f(\fvec{q}, t)$ gives the complex-valued weights of this quantum
mixture of states. It should belong to the space of square-integrable functions
that we note here $L^2(E)$. The expectation value of any observable $\hat{O}$
for a GHW state has the compact form
\begin{equation}
  \qavg{\op{O}}(t) =
    \iint\diff{\fvec{q}}\diff{\fvec{q}'}
      f^\star(\fvec{q}, t)
      \kerRep{O}(\fvec{q}, \fvec{q}')
      f(\fvec{q}', t).
\end{equation}
We used here the notation $\kerRep{O}(\fvec{q}, \fvec{q}')$ for the kernel of
the observable defined by
\begin{equation}
  \kerRep{O}(\fvec{q}, \fvec{q}') = \bkHFB[\op{O}]{\fvec{q}}{\fvec{q}'}.
\end{equation}
Significant kernels that we will discuss through this review are the norm
kernel and the energy (or Hamiltonian) kernel. They are defined as
\begin{align}
\label{eq:gcm:nhker}
  \hker(\fvec{q}, \fvec{q}')
    &= \bkHFB[\op{H}]{\fvec{q}}{\fvec{q}'}
    & \text{(Hamiltonian),} \\
  \nker(\fvec{q}, \fvec{q}')
    &= \bkHFB[\opId]{\fvec{q}}{\fvec{q}'}
    & \text{(norm).}
\end{align}

We emphasize that the choice of collective coordinates $\fvec{q}$ is somehow
arbitrary. From one choice of collective coordinate, we may switch to a
different one while keeping invariant the space of GHW states. We can show this
by defining a change of variable $\varphi$
\begin{equation}\label{eq:gcm:metric}
  \fvec{a} = \varphi(\fvec{q}).
\end{equation}
Then we may consider the GHW ansatz built on the transformed generator states
$\kmHFB{\fvec{a}} = \kHFB{\varphi^{-1}(\fvec{a})}$
\begin{equation}
\label{eq:gcm:GHWgenmetric}
  \ket{\tilde{\psi}(t)} =
    \int_{\fvec{a}\in \varphi(E)}
    \diff{\fvec{a}}
      \kmHFB{\fvec{a}}
      \tilde{f}(\fvec{a}, t).
\end{equation}
Any GHW state defined by Eq.~\eqref{eq:gcm:GHW} can be cast into
Eq.~\eqref{eq:gcm:GHWgenmetric} with the weight function
\begin{align}
  \tilde{f}(\fvec{a}, t)
    &= f(\varphi^{-1}(\fvec{a}), t)
       |\det(J_{\varphi}(\fvec{a}))|^{-1}.
\end{align}
Here $J_{\varphi}$ is the Jacobian matrix of the coordinate transformation.
Also, the formula for the expectation value observables is invariant by this
change of coordinate. Typically we have in the $\fvec{a}$ representation
\begin{equation}
  \qavg{\op{O}}(t) =
    \iint\diff{\fvec{a}} \diff{\fvec{a}'}
      \tilde{f}^\star(\fvec{a}, t)
      \kerRep{O}(\fvec{a}, \fvec{a}')
      \tilde{f}(\fvec{a}', t),
\end{equation}
with 
\begin{equation}
  \kerRep{O}(\fvec{a}, \fvec{a}') =
    \bmHFB{\fvec{a}}\op{O}\kmHFB{\fvec{a}'}.
\end{equation}
Although applying such a change of variable does not change the physics of the
ansatz, it does change intermediate quantities involved in the GCM framework.
In some cases, it may be essential to change the variables to obtain valuable
mathematical properties of the kernel
operators~\cite{deKinematics1978,ringNuclear2004}.

As a final remark, we would like to highlight that the integral of
Eq.~\eqref{eq:gcm:GHW} may not be well defined for some weight functions and
family of generator states. The Ref.~\cite{deKinematics1978} gives a
mathematically rigorous presentation of the GCM framework. We retain from this
work that a sufficient condition for the GHW ansatz to be valid is that norm
kernel defines a bounded linear operator on $L^2(E)$.  

\subsection{Griffin-Hill-Wheeler equation}
\label{subsec:gcm:HWeq}

The time-dependent Schr\"odinger equation in the entire many-body Hilbert space,
\begin{equation}\label{eq:gcm:schrodinger}
  \left(\op{H} - i\hbar\frac{{\rm d}}{{\rm d}t}\right)\kTDGCM{t} = 0,
\end{equation}
drives the exact time evolution of a many-body system $\kTDGCM{t}$. We assume
here that all the interactions between the nucleons are encoded into the
Hamiltonian $\op{H}$ acting on the full many-body space. From this starting
point, the TDGCM equation of motion can be obtained by assuming that at any time $t$:
\begin{enumerate}
  \item the wave function of the system keeps the form of Eq.~\eqref{eq:gcm:GHW},
  \item the equality
    \begin{equation}\label{eq:gcm:projschrodinger}
      \bra{\Phi} 
        \left(\op{H}-i\hbar\frac{{\rm d}}{{\rm dt}}\right)
      \kTDGCM{t} = 0
    \end{equation}
    is satisfied for every GHW state $\ket{\Phi}$.
\end{enumerate}
In other words, we impose that the residual
$(\op{H}-i\hbar {\rm d}/ {\rm dt})\kTDGCM{t}$ is orthogonal to the space of GHW
states. This last assumption is equivalent to a Frenkel's variational principle
whose link to other time-dependent variational principles is discussed in
Ref.~\cite{lowdinComments1972}. By injecting the GHW ansatz~\eqref{eq:gcm:GHW}
into~\eqref{eq:gcm:projschrodinger}, we obtain
\begin{equation}\label{eq:gcm:integralHW}
  \int\int \diff{\fvec{q}}\diff{\fvec{q'}}
    f^{\star}_{\Phi}(\fvec{q'})
      \Big(
        \hker(\fvec{q'}, \fvec{q})
      - i\hbar\nker(\fvec{q'}, \fvec{q})\frac{{\rm d}}{{\rm dt}}
      \Big)
      f(\fvec{q}, t) = 0.
\end{equation}
Here $f_{\Phi}$ is the mixing function defining the GHW state $\ket{\Phi}$.
Solving Eq.~\eqref{eq:gcm:integralHW} for any state $\ket{\Phi}$ is equivalent
to look for a function $f$ verifying the so-called Griffin-Hill-Wheeler
equation in its time-dependent form
\begin{equation}\label{eq:gcm:stdHW}
  \forall \fvec{q'}: \quad 
  \int\diff{\fvec{q}}
    \Big(
        \hker(\fvec{q'}, \fvec{q}) \\
      - i\hbar\nker(\fvec{q'}, \fvec{q})\frac{{\rm d}}{{\rm dt}}
      \Big)
      f(\fvec{q}, t) = 0.
\end{equation}

The time-evolution of the norm and the energy reads
\begin{align}
  \frac{\rm d}{{\rm d}t}\bkTDGCM{t}{t}
    &= \frac{i}{\hbar} \bkTDGCM[(\op{H}^\dagger-\op{H})]{t}{t}, \\
  \frac{\rm d}{{\rm d}t}E(t)
    &= \frac{i}{\hbar} \bkTDGCM[(\op{H}^\dagger-\op{H})\op{H}]{t}{t}.
\end{align}
Thus, this equation of motion preserves the norm of the wave function and the
total energy of the system if the many-body Hamiltonian is Hermitian. However,
it is not always the case. To simulate open systems, for instance in the context
of nuclear reactions, a common practice consists in adding an imaginary absorption
term to the Hamiltonian that acts in the neighborhood of the finite simulation
box.
Finally, the time-dependent GHW equation is
a continuous system of integrodifferential equations. Its non-local nature in
the $\fvec{q}$ representation brings a serious hurdle to its numerical solving.

\subsection{Mapping to the collective wave functions}
\label{subsec:collobs}

The equation of motion~\eqref{eq:gcm:integralHW} and an initial condition
for the system is sufficient to determine the dynamics in the TDGCM framework.
It is possible to numerically integrate in time this equation with an implicit
scheme such as Crank-Nicolson~\cite{crankPractical1996}. However, the TDGCM
framework offers another natural approach that turns out to be both
enlightening from the mathematical perspective and more stable from a numerical
point of view. This method resorts on a mapping between the GHW states and some
functions of the collective coordinate $\fvec{q}$. The rigorous mathematical
construction of this mapping in a general case is detailed in
Ref.~\cite{deKinematics1978}. Here we will only build this mapping in the case
where the norm kernel $\nker$ is of Hilbert-Schmidt
type~\cite{reedMethods1972}. It is the case as long as the domain $E$ of the
collective coordinates is bounded, which is valid for a wide range of applications.

To start with, we recall that any kernel $\kerRep{O}(\fvec{q},\fvec{q'})$ also
defines a linear operator acting on the space of functions $L^2(E)$ 
\begin{equation}
  (\opRep{O}f)(\fvec{q}) =
    \int_{\fvec{q'}\in E} \diff{\fvec{q'}}
      \kerRep{O}(\fvec{q},\fvec{q'})
      f(\fvec{q'}),
\end{equation}
as long as this integral is mathematically defined.
The Hilbert-Schmidt property of the norm operator implies the existence of a
complete, discrete and orthonormal family of functions $\{ u_i(\fvec{q}) \}_i$
of $L^2(E)$ that diagonalizes the linear operator associated with the norm
kernel
\begin{equation}
 \forall i>0: \quad \nker u_i = \lambda_i u_i.
\end{equation}
Since $\nop$ is a Hermitian positive semidefinite operator, its eigenvalues are
real and positives. We adopt here the convention where they are sorted by
decreasing order and assume that only the first $r$ eigenvalues are not zero.
From this diagonalization, we can split the space of functions $f$ into two
orthogonal subspaces: the one associated with the vanishing eigenvalues and the
one associated with the strictly positive eigenvalues. Formally, we write down
the two projectors
\begin{align}
 \kerRep{Q}({\fvec{q}, \fvec{q'}}) &=
   \sum_{i \le r} u_i(\fvec{q}) u_i^{\star}(\fvec{q'}) \\
 \kerRep{P}({\fvec{q}, \fvec{q'}}) &=
   \sum_{i > r}   u_i(\fvec{q}) u_i^{\star}(\fvec{q'})
\end{align}
with
\begin{equation}
 \opRep{Q} + \opRep{P} = \mathbbm{1}_{L^2(E)}.
\end{equation}
The projected space $\opRep{P}L^2(E)$ is associated with the null eigenvalues
of the norm operator $\nop$. Any GHW state built from a weight function
belonging to this space gives the null many-body state. Its orthogonal
complement is the subspace $Q(E) = \opRep{Q} L^2(E)$. We call collective wave
functions, the functions living in this subspace.

We can define uniquely the positive hermitian square-root of $\nker$ (which is
also Hermitian) with
\begin{equation}
 \nker(\fvec{q},\fvec{q'}) 
 = \int_{\fvec{a}\in E} 
  \nker^{1/2}(\fvec{q},\fvec{a})
  \nker^{1/2}(\fvec{a},\fvec{q'}) \diff{\fvec{a}}.
\end{equation}
We can, therefore, associate to any GHW state its collective wave function
$g(\fvec{q})\in Q(E)$ by the equation
\begin{equation}
 g = \nker^{1/2} f.
\end{equation}
Conversely, the operator $\nker^{1/2}$ is invertible in $Q(E)$. Therefore for
any collective wave function $g\in Q(E)$, one can build its corresponding GHW
state with the weight function 
\begin{equation}\label{eq:gcm:ffromg}
 f = \nker^{-1/2} g.
\end{equation}
Finally, this mapping between $Q(E)$ and the GHW states is isometric as we may
show that for any pair of GHW states $\Psi$ and $\Phi$ we have the property
\begin{align}
  \braket{\Psi}{\Phi} 
  &= \braket{g_{\Psi}}{g_{\Phi}} \nonumber \\
  &= \int_{\fvec{q}\in E} g_{\Psi}^{\star}(\fvec{q}) g_{\Phi}(\fvec{q}) \diff{\fvec{q}}.
\end{align}

Going further, any many-body observable $\op{O}$ can be mapped into a collective
operator $\col{O}$ acting on the space $Q(E)$. This operator is defined by
\footnote{Note that such a definition is possible for any observable $\op{O}$
due to the property
\begin{equation*}
 \opRep{Q}\opRep{O}=\opRep{O}.
\end{equation*}
}\begin{equation}\label{eq:colop}
 \col{O} = \nker^{-1/2} \opRep{O} \nker^{-1/2}.
\end{equation}
The isometry of the mapping gives a simple mean to compute matrix elements of
observables.
\begin{equation}\label{eq:gcm:opmapping}
  \braket[\hat{O}]{\Psi}{\Phi}
  = \braket[\col{O}]{g_{\Psi}}{g_{\Phi}}
\end{equation}

Finally, we can reduce the TDGCM equation of motion (Eq.~\eqref{eq:gcm:stdHW})
in this language. It becomes a time-dependent Schr\"odinger equation for the
collective wave function
\begin{equation}
\label{eq:tdgcm_coll}
 i\hbar \dot{g} = \col{H} g.
\end{equation}
This equation of motion presents several practical advantages compared to
Eq.~\eqref{eq:gcm:stdHW}.
The collective Hamiltonian $\col{H}$ is, in general, still non-local, but the
time derivative of $g$ has an explicit expression. It opens the possibility
of using faster time integration schemes at the cost of computing first the
collective Hamiltonian through Eq.~\eqref{eq:colop}. Also, the collective wave
function is expected to have a smoother behavior compared to the weight
function $f$. This comes directly from Eq.~\eqref{eq:gcm:ffromg} where we see
that eigenvalues of the norm kernel approaching zero add diverging components
to $f$. The equation~\eqref{eq:tdgcm_coll} may be directly solved by
discretizing the collective wave function $g(\fvec{q})$. In many cases, it is
appropriate to solve it directly in the representation given by the basis
$\{u_i(\fvec{q})\}_{i\le r}$. The collective Hamiltonian $\col{H}$, as well as
other collective observables, are indeed easier to compute in this particular
basis.

\subsection{Difficulties related to the energy kernel}
\label{sec:multireference}

We discussed general features of the TDGCM approach valid for any family of
generator states. In nuclear physics, most applications of the GCM rely on
families of Bogoliubov vacua. A crux of the GCM approach is then the
determination of the norm and Hamiltonian kernels between such many-body
states. The Ref.~\cite{robledoSign2009} provides a general and now-standard
approach to fully determine the norm kernel between Bogoliubov vacua based on
the calculation of a matrix Pfaffian. However, the evaluation of the
energy kernel in nuclear physics applications suffers from several major
difficulties. The origin of these flaws
stems from the fact that our practical applications do not rely on a linear
many-body Hamiltonian but some effective Hamiltonians or energy density
functionals. This topic was extensively discussed in the context of static GCM
for nuclear structure~\cite{anguianoParticle2001,lacroixConfiguration2009,%
benderParticlenumber2009,duguetParticlenumber2009,sheikhSymmetry2019}. We
briefly list here the pitfalls raised by the determinations of the energy
kernel in practical nuclear applications.

\subsubsection{Neglecting some exchange terms}
A common practice to avoid unbearable numerical costs is the neglection or the
approximation of parts of the many-body Hamiltonian. For instance, it is
widespread to use the Slater approximation of the Coulomb exchange term or to
neglect the exchange part of the pairing force between
nucleons~\cite{ryssensSolution2015}. Although convenient from a numerical point
of view, it was shown in Ref.~\cite{donauCanonical1998} that such approximations
may introduce poles in the expression of the energy kernel. These poles lead to
a divergence when calculated between some Bogoliubov vacua. The
Refs.~\cite{anguianoCoulomb2001,almehedPairing2001} illustrate this behavior in
a case of particle number symmetry restoration.

\subsubsection{Violation of symmetries by energy density functionals}
In many practical applications, the nucleon-nucleon interaction is encoded in
an energy density functional (EDF). Using such a formalism in combination with
a GCM mixture of states requires a sound definition of a multireference energy
density functional~\cite{lacroixConfiguration2009}. Such a definition is often
provided and implemented in the form of the reduced energy kernel
$h(\fvec{q},\fvec{q'}) = \hker(\fvec{q},\fvec{q'}) / \nker(\fvec{q},\fvec{q'})$
between two non-orthogonal Bogoliubov vacua. For a two-body Hamiltonian case,
the reduced energy kernels may be expressed from the generalized Wick theorem
\begin{equation}
\label{eq:gcm:hgwt}
  h(\fvec{q,q'}) =
    \sum_{ij}
      t_{ij} \rho^{\fvec{qq'}}_{ji}
  + \frac{1}{2}
    \sum_{ijkl}
      \bar{v}_{ijkl} \rho_{ki}^{\fvec{qq'}} \rho^{\fvec{qq'}}_{lj}
  + \frac{1}{4}
    \sum_{ijkl}
      \bar{v}_{ijkl}
      \kappa_{ij}^{\fvec{qq'}*}
      \kappa^{\fvec{qq'}}_{kl}.
\end{equation}
It involves the matrix elements of the one- and two-body parts of the
interaction $t$ and $\bar{v}$ as well as transition densities such as
\begin{equation}
\label{eq:gcm:mixed}
  \rho^{\fvec{qq'}}_{ij}
  = \frac{\bkHFB[\cPw{j}\aPw{i}]{\fvec{q}}{\fvec{q'}}}
         {\bkHFB{\fvec{q}}{\fvec{q'}}}.
\end{equation}
In the practical implementations of the multireference EDF approach, such a
kernel is defined by analogy as the same bilinear form whose coefficients
come from a fit procedure. The main differences compared to the EDF case are:
\begin{enumerate}
 \item the coefficients defining the EDF may depend on some densities of the
   system,
 \item the coefficients in the particle-particle channels may differ from the
   ones in the particle-hole channels,
 \item the matrix $\bar{v}$ may not be antisymmetric.
\end{enumerate}
As detailed in Refs.~\cite{lacroixConfiguration2009,
dobaczewskiParticlenumber2007}, the violation of these properties leads in some
cases to a divergence of the reduced energy kernel that biases or prevents
practical applications.

\subsubsection{Density dependent terms of energy density functionals}

In an EDF framework, the coefficients of Eq.~\eqref{eq:gcm:hgwt} depend on the
density of the system. The exact formulation of this dependency is yet subject
to an arbitrary choice, especially for the non-diagonal part of the kernel.
Several prescriptions have been developed and tested during the last two
decades~\cite{duguetDensity2003,robledoRemarks2010}. A prescription that
fulfills many important conditions expected from a Hamiltonian is the
transition density defined by Eq.~\eqref{eq:gcm:mixed} (see
Ref.~\cite{rodriguez-guzmanCorrelations2002}). However, this prescription
yields to complex-valued densities. It is then incompatible with most of the
EDFs developed at the mean-field level with terms that contain a non-integer
power of the density. Finding a satisfying pair of density prescription and EDF
valid for GCM calculations is still an open problem.

In conclusion, the current usage of the GCM formalism with effective
Hamiltonian or energy density functionals suffers from several formal and
practical flows when it comes to determining the energy kernel. This situation
has been a major obstacle to the development of GCM applications in nuclear
physics in the last years. Several ongoing efforts attempt to overcome this
difficulty by building new energy functionals valid for multireference
calculations~\cite{bennaceurNonlocal2017} or going toward \textit{ab initio}
treatments~\cite{yaoInitio2019}.

\subsection{Fission dynamics with the exact TDGCM}
\label{sec:tdgcm_appli}

The exact solving of the time-dependent GHW equation in a realistic case has
rarely been carried out. To our knowledge, the only published work tackling
this task is presented in~\cite{verriereDescription2017,verriereFission2017} in
the context of fission. It shows the challenges raised by an exact TDGCM
calculation, especially when dealing with large collective coordinate domains.

In Ref.~\cite{verriereDescription2017}, the authors used the TDGCM to describe
the reaction \textsuperscript{239}Pu(n,f). This study relies on two common
collective coordinates for fission, namely $q_{20}$ and $q_{30}$, that are
associated with the expectation value of the quadrupole and the octupole
moments of the one-body density. The dynamics in this collective space accounts
for the evolution from a compound to a fragmented system with, also,
information on the mass asymmetry between the two fragments produced. It is
well suited to determine the mass yields of the fragments. The set of
constrained HFB solutions (a total of 20212) obtained for a wide range of these
collective coordinates forms the family of generator states. Each generator
state is practically obtained with a finite-range Gogny interaction in its D1S
parametrization. A two-center axial harmonic oscillator basis with 12 shells
has been used where the parameters defining the basis have been optimized for
each value of the collective coordinates.

\begin{figure}[ht]
  \includegraphics[width=0.99\linewidth]{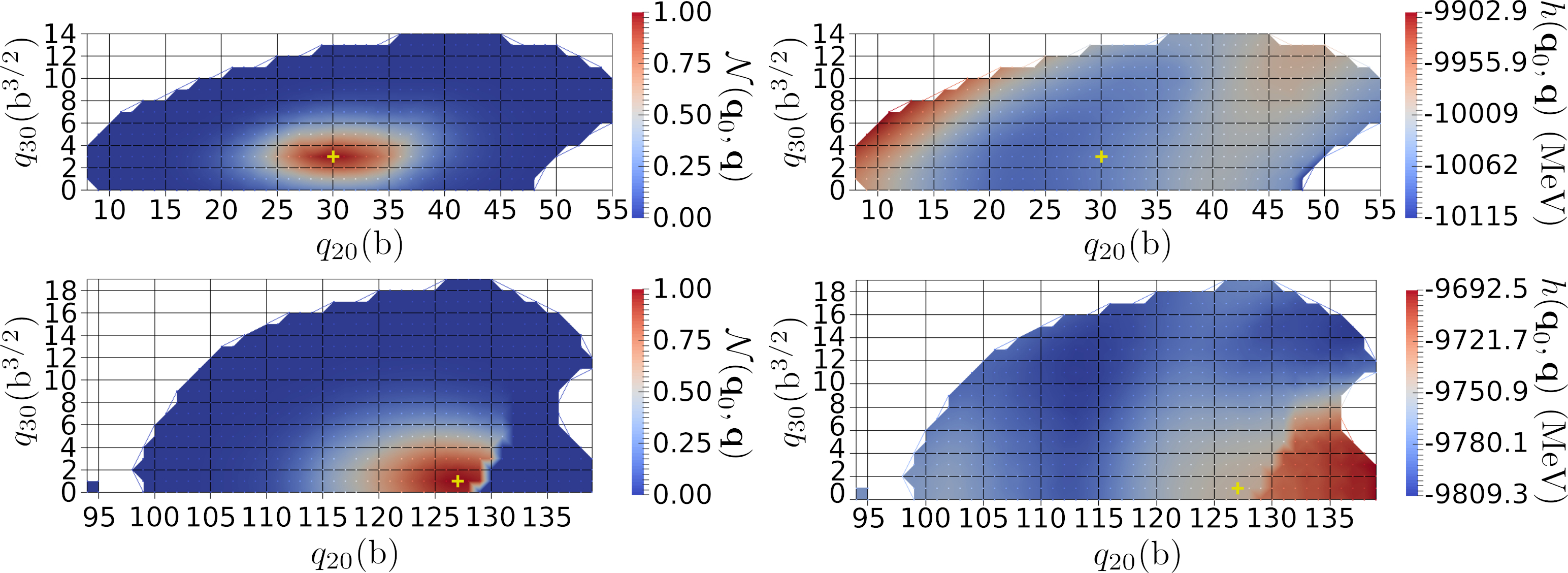}
  \caption{
    Left panels:
    Overlaps $\nker(\fvec{q}_0, \fvec{q})$ as a function of $\fvec{q}$ and
    where $\fvec{q}_0=(30,3)$, in barn units, corresponds to the ground state
    (upper panel) or $\fvec{q}_0=(127,1)$, in barn units, which corresponds to
    a point at higher elongation (lower panel). This was obtained for a
    \textsuperscript{240}Pu nucleus. The white parts correspond to values below
    a threshold of $1.0\times 10^{-4}$. The yellow crosses correspond to
    $\fvec{q}_0$.
    Right panels: same as the corresponding left panels for the reduced energy
    kernel $h(\fvec{q}_0, \fvec{q})$.
  }
  \label{fig:GCM:240PuN}
\end{figure}

\begin{figure}[ht]
\centering
\includegraphics[width=0.80\linewidth]{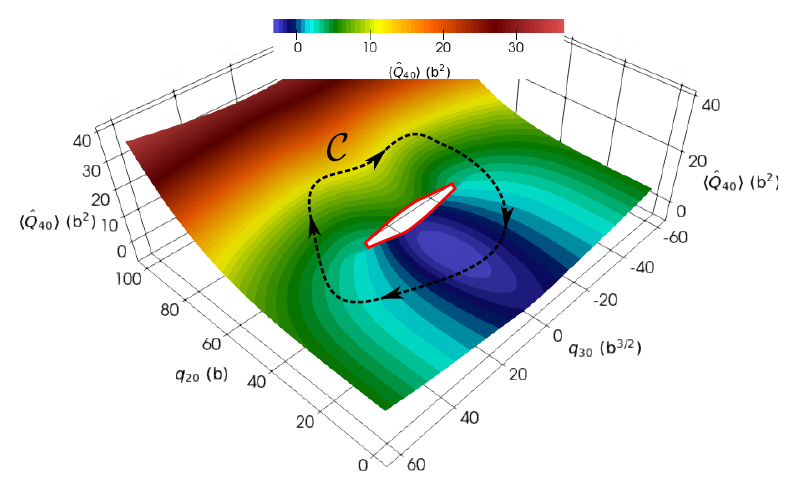}
  \caption{Schematic representation of a discontinuity in two-dimensional
  calculations with constraints on $q_{20} = \langle\hat{Q}_{20}\rangle$
  (x-axis) and $q_{30} = \langle\hat{Q}_{30}\rangle$ (y-axis). The z-axis
  and the color scale are associated with the average values of the
  unconstrained hexadecapole moment $\langle\hat{Q}_{40}\rangle$.
  }
  \label{fig:GCM:disc}
\end{figure}

The norm kernel has been calculated for each couple of generator states. The
upper-left panel of Fig.~\ref{fig:GCM:240PuN} presents its values between the
mean-field ground-state and the surrounding points, whereas the lower-left
panel of the figure shows its values obtained for a more elongated
configuration in the potential energy surface (PES). We see that the overlaps
are above $\epsilon_{\rm thresh}=1.0\times 10^{-4}$ only in a neighborhood of
$\fvec{q}_0$ in both cases. As noted in~\cite{ringNuclear2004}, it is due to
the large number of nucleons in the system. 
This behavior is at the heart of the Gaussian Overlap Approximation, discussed
in more detail in Sec.~\ref{sec:goa}. The reduced Hamiltonian $h(\fvec{q}_0,
\fvec{q})$, defined as the ratio between the collective Hamiltonian and the
norm kernel
\begin{equation}
  h(\fvec{q}_0, \fvec{q}) =
    \frac{\bkHFB[\hat{H}]{\fvec{q}_0}{\fvec{q}}}{\bkHFB{\fvec{q}_0}{\fvec{q}}},
\end{equation}
has also been calculated for all overlaps greater than $\epsilon_{\rm thresh}$.
In this work, only the kinetic and central terms of the interaction were
included. The right panels of Fig.~\ref{fig:GCM:240PuN} presents the slices of
the reduced Hamiltonian for the same cases as in its left panels. The relative
variation of the reduced Hamiltonian (where the norm kernel above the
threshold) is almost constant, being only 2\% around the ground state and 1\%
for the elongated configuration. In addition to the overlaps rapid decrease
discussed above, it numerically justifies the standard second-degree polynomial
approximation of this quantity (a further study with all the terms of the
interaction is, however, required). The bottom panels highlight a discontinuous
behavior around $q_{20}\approx 130$ b. This specific discontinuity is due to
the existence of two competing valleys in the three-dimensional PES obtained by
adding the hexadecapole moment $q_{40} = \langle\op{Q}_{40}\rangle$ as a
collective DoF~\cite{dubrayNumerical2012}. Such a discontinuity gives a similar
label in the collective space to two HFB states that are far in the full
many-body space. The Fig.~\ref{fig:GCM:disc} is an illustration of such a
discontinuity in a two-dimensional PES embedded in a three-dimensional
collective space. It is not possible to reduce the loop $\mathcal{C}$ to a
point: the discontinuity is a hole whose edges are highlighted by
the red line of Fig.~\ref{fig:GCM:disc}. Such a discontinuity may add spurious
boundary effects in the description of the reaction of interest. It is
especially the case when the discontinuity appears in an area of the collective
space that gives important contributions to the targeted observables. Note that
in approximate treatments such as the ones based on the Gaussian Overlap
Approximation, discontinuities are always neglected, leading to a spurious
connection between distant regions of the full many-body space.

\begin{figure}[ht]
 \centering
  \includegraphics[width=0.7\linewidth]{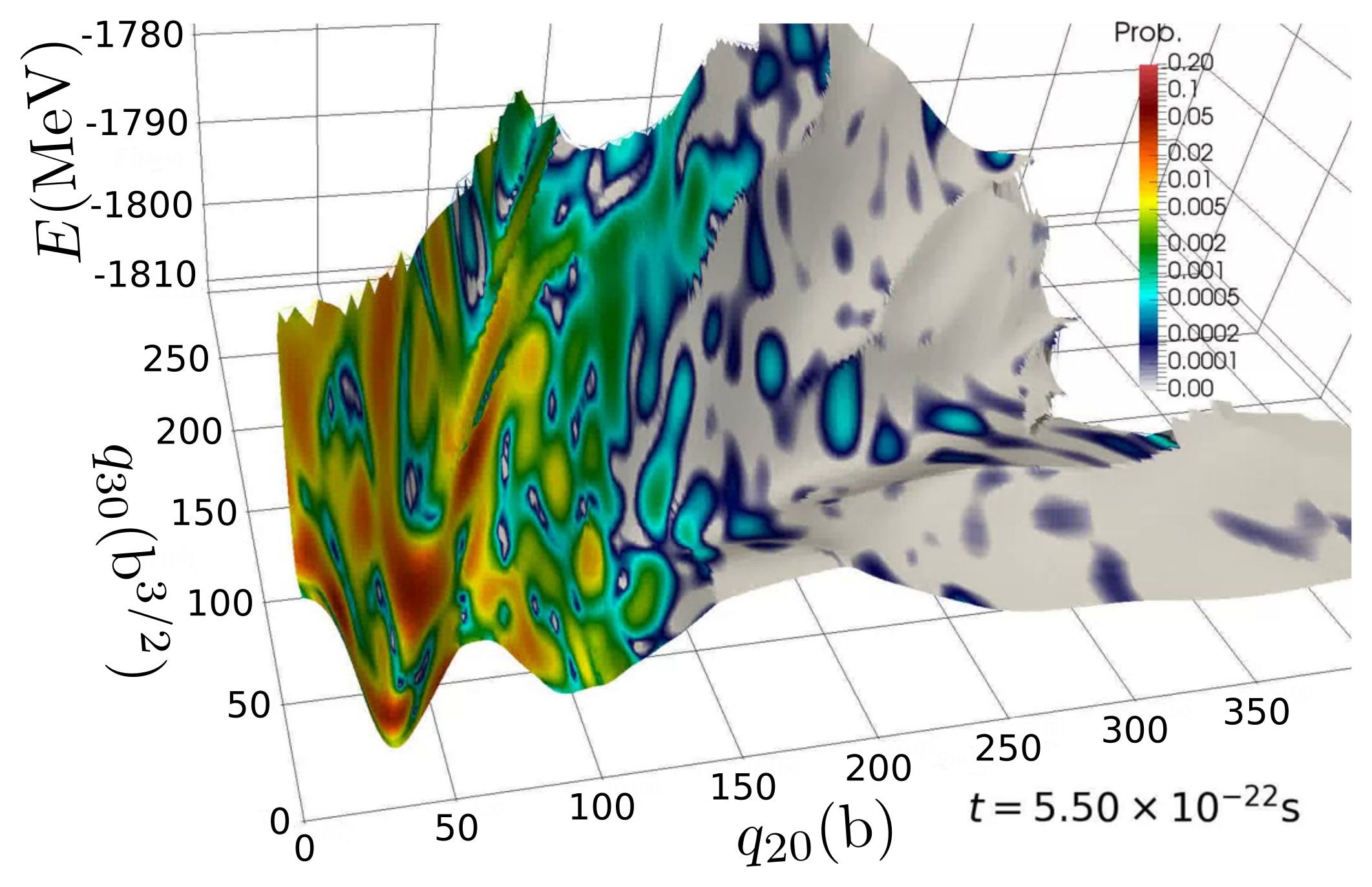}
  \caption{
    The gray surface represents the generator states' HFB energy as a
    function of $q_{20}$ and $q_{30}$. On top of this, the color map gives the
    quantity $\mathbb{P}(\fvec{q}, t=0.550\text{ zs})$ in the same conditions
    than those of Fig.~\ref{fig:GCM:240PuN}.\label{fig:GCM:240PuT}
  }
\end{figure}

It is possible to determine the time evolution of the weight function
$f(\fvec{q},t)$ of the GHW ansatz~\eqref{eq:gcm:GHW}. In cases where the size
of the discretized space of the collective coordinate is still tractable, this
task has been achieved through a direct diagonalization of the collective
Hamiltonian~\cite{verriereFission2017}. For this two-dimensional application,
the straightforward diagonalization involves a prohibitive numerical cost. It
is still possible to use a Crank-Nicolson method to integrate in time the GHW
equation~\eqref{eq:gcm:stdHW}. Figure~\ref{fig:GCM:240PuT} presents a snapshot
at time $t=0.55$ zs of the quantity $\mathbb{P}(\fvec{q}, t)$ defined as
\begin{equation}
  \mathbb{P}(\fvec{q}, t) \equiv
    \bTDGCM{t}
    \left(
      \kHFB{\fvec{q}}\bHFB{\fvec{q}}
    \right)\kTDGCM{t}.
\end{equation}
This corresponds to the probability to measure the system in the state
$\kHFB{\fvec{q}}$ \footnote{
  Note that due to the non-orthogonality of the generator states its sum over
  all the points $\fvec{q}$ is not equal to 1.
}. Even though the simulation was a proof-of-concept, we see that the bottom of
the asymmetric valley is slightly more populated than the other parts of the
PES near scission. This leads mostly to asymmetric fission fragments, which is
in agreement with experimental data~\cite{nishioMeasurement1995,
tsuchiyaSimultaneous2000}. The GCM wavefunction evolves in a slightly non-local
way in the collective space (in the range of the width of the overlaps along
$\fvec{q}-\fvec{q}'$), leading to non-zero probability ``drops'' appearing and
disappearing along the time-evolution of the system.
The most time-consuming part was the calculation of the norm and Hamiltonian
kernels that required the use of 512 cpus for two weeks ($\sim 170,000$
cpu.h). The calculation of the time-evolution of the weight function $f(\fvec{q},t)$
for times up to 0.55zs was done using 64 cpus for one week ($\sim 10,000$
cpu.h). The short length of time for which the weight function was determined
is not enough for the calculation of mass and charge probability distributions.
A more realistic calculation would require at least $200,000$ cpu.h, for the
determination of the weight function up to 10zs only.

The principal difficulty of such an application stems from the big size of the
discretized space of the collective coordinates (substantially bigger, for
example, than in the case of the static GCM calculations for nuclear
structure). This makes the computation of the norm and Hamiltonian kernel
intensive but still embarrassingly parallel. Besides, in the case of fission,
techniques to determine the post-scission observables of the fragments still
need to be developed for the exact TDGCM.  For instance, some simplifying
hypotheses on the way to treat open domains of collective coordinates are
commonly used under the Gaussian Overlap
Approximation~\cite{goutteMicroscopic2005} but are no longer valid in the exact
TDGCM framework.
 
\section{Gaussian Overlap Approximation (GOA)}
\label{sec:goa}

In its straightforward application, the TDGCM leads to a non-local equation of
motion that must be solved in a high-dimensional space in most of the practical
calculations. As mentioned in Sec.~\ref{sec:gcm}, solving this equation
involves a high numerical cost that strongly hurdles its applications in
nuclear physics. Several approximate treatments of the TDGCM have been
developed with the aim to build a local equation of motion for the collective
wave function $g(\fvec{q},t)$ (cf. Eq.~\eqref{eq:tdgcm_coll}). The Gaussian
overlap approximation (GOA) is one of these approximations, which leverages the
fact that the overlap and Hamiltonian kernels can, in some cases, be
parametrized in terms of Gaussians of the variable $\fvec{q}$. In its static
form, the GOA has been largely used and applied for nuclear structure.
Especially, it provides a nice bridge between the Bohr Hamiltonian equation
that was first formulated in~\cite{bohrCoupling1952} and a quantum treatment
based on the $3A+A$ nucleons degrees of freedom~\cite{kumarComplete1967,
libertMicroscopic1999,delarocheStructure2010,fuBeyond2013,
matsuyanagiMicroscopic2016}. Extensive reviews of the static version of the GOA
can be found in Refs.~\cite{reinhardGeneratorcoordinate1987,ringNuclear2004}.
We focus here on its time-dependent flavor.

\subsection{TDGCM+GOA with time even generator states}
\label{sec:std_goa}

\subsubsection{Main assumptions}
\label{sec:goa_assumptions}
In its most standard form, the GOA framework assumes the following situation:
\begin{enumerate}
  \item we have a family of normed generator states $\{\kHFB{\fvec{q}}\}$
    parametrized by a vector of real coordinates $\fvec{q} \in \Re^{\Qdim}$;
  \item all the states of the set are time-even, \textit{i.e}, they are their
    own symmetric by the time-reversal operation;
  \item the function $\fvec{q} \rightarrow \kHFB{\fvec{q}}$ is continuous and twice
    derivable;
  \item the overlap between two arbitrary generator states can be approximated
    by a Gaussian shape
    \begin{equation}
      \overlap(\fvec{q}, \fvec{q}') \simeq
        \operatorname{exp}
          \left[
            -\frac{1}{2}(\fvec{q}-\fvec{q'})^t G(\bar{\fvec{q}})
            (\fvec{q}-\fvec{q'})
          \right],
    \end{equation}
    with $\bar{\fvec{q}}= (\fvec{q}+\fvec{q'}) / 2$ and $G(\bar{\fvec{q}})$ a real
    positive definite matrix;
  \item the Hamiltonian kernel can be approximated by
    \begin{equation}
      \hker(\fvec{q}, \fvec{q'}) \simeq
        \nker(\fvec{q}, \fvec{q'}) h(\fvec{q}, \fvec{q'}),
    \end{equation}
    where $h(\fvec{q}, \fvec{q'})$, a polynomial of degree two in the collective
    variables $\fvec{q}$ and $\fvec{q'}$, is the reduced Hamiltonian.
\end{enumerate}
In most applications of the TDGCM+GOA, the generator states are built as
constrained Hartree-Fock-Bogoliubov states of even-even nuclei which ensures
the time even property. The question is then: what are the situations where
the Gaussian shape approximation is verified within a small error? Already from
the time-reversal symmetry, we can infer that the overlaps are real and
symmetric in $(\fvec{q} - \fvec{q'})$. Therefore, the following relation is
satisfied in the vicinity of $\fvec{q}$
\begin{equation}
  \bkHFB{ \fvec{q} + \frac{\fvec{s}}{2} }{ \fvec{q}-\frac{\fvec{s}}{2} } =
    \operatorname{exp}\left[
      \operatorname{ln}\left(
        \braket{\fvec{q}+\frac{\fvec{s}}{2}}{\fvec{q}-\frac{\fvec{s}}{2}}
      \right)
    \right].
\end{equation}
A Taylor development of this expression up to order two in $\fvec{s}$ already
yields locally a Gaussian shape without any additional assumption
\begin{equation}
  \bkHFB{ \fvec{q} + \frac{\fvec{s}}{2} }{ \fvec{q}-\frac{\fvec{s}}{2} } 
  = \operatorname{exp}
  \left[
    -\frac{1}{2} \fvec{s}^t G(\fvec{q}) \fvec{s} 
                        + o(\fvec{s}^2)
  \right]
\end{equation}
with
\begin{align}
  G_{ij}(\fvec{q}) &= \braket{\partial_i\HFB(\fvec{q})}{\partial_j\HFB(\fvec{q})},\\
  \ket{\partial_i \HFB (\fvec{q})} &= 
    \left.
      \frac{\partial \ket{\HFB(\fvec{a})} }{\partial \fvec{a}_i}
    \right|_{\fvec{q}}.
\end{align}
We used here some identities coming from the fact that generator states are
normalized. In situations where the coordinates correspond to some collective
deformations of the nucleus, it turns out that the Gaussian shape holds for
larger values of $\fvec{s}$. This is justified from the central limit theorem
in Ref.~\cite{reinhardGeneratorcoordinate1987} for Slater determinants or in
Ref.~\cite{ringNuclear2004} for Bogoliubov vacua. It especially holds for heavy
nuclei.

Finally, note that although we limit here our description to the case of
time-even generator states, it is possible to build a GOA framework without
assuming this symmetry. Such a generalization can be found, for instance, in
Ref.~\cite{reinhardGeneratorcoordinate1987}.

\subsubsection{Equation of motion}
\label{sec:goa_eom}

Starting from the GOA hypothesis, one can reduce the equation of
motion~\eqref{eq:tdgcm_coll} to a local equation involving the first and
second-order derivatives of the collective wave function. In this section, we
give only the main ideas to derive this local equation. For more exhaustive
demonstrations, we refer the reader to
Refs.~\cite{reinhardGeneratorcoordinate1987,ringNuclear2004,krappeTheory2012}.

In its historical version, the GOA framework assumes that the width of the
Gaussian shape is constant. However, in most of the practical cases, this assumption
is too restrictive. To overcome this issue, a series of papers published in the
70-80's generalized the GOA framework to account for a varying Gaussian
width~\cite{onishiLocal1975,gozdzExtended1985,krappeTheory2012}. The idea is to
perform a change of collective variables to recover the constant width case.
The mapping between the new collective coordinates $\fvec{\alpha}$ and the
original ones $\fvec{q}$ reads 
\begin{equation}\label{eq:goa_change_var}
  \fvec{\alpha}(\fvec{q}) =
    \int_{\fvec{a}\in C_{0}^{\fvec{q}}}  G^{\frac{1}{2}}(\fvec{a}) \diff{\fvec{a}}
\end{equation}
where $C_0^{\fvec{q}}$ is a path from the origin to $\fvec{q}$. With this new
labeling of the generator states, we get
\footnote{
  Note that this assumes (i) that the integrals of $G^{1/2}(\fvec{a})$
  are independent of the integration path (ii) that its evaluation properly
  approximates the average of $G^{1/2}$ on the path at the central point of the
  path~\cite{gozdzExtended1985}.
}
\begin{equation}
  \bkHFB{\fvec{\alpha}}{\fvec{\alpha'}}
  \simeq
    \operatorname{exp}\left[
      -\frac{1}{2}(\fvec{\alpha} - \fvec{\alpha}')^2
    \right].
\end{equation}
We can therefore perform all the derivations with the $\fvec{\alpha}$
coordinates and make the inverse transformation on the final expressions only.

Starting with this simple form of the overlap, we seek an equation of motion
involving a local collective Hamiltonian in the collective coordinate
representation. The Gaussian shape of the norm kernel allows expressing its
positive Hermitian square root analytically as
\begin{equation}
  \overlap^{1/2}(\fvec{\alpha}, \fvec{\alpha}') =
    C \cdot \operatorname{exp} \left[ 
              -(\fvec{\alpha} - \fvec{\alpha}')^2
            \right],
\end{equation}
where the constant $C$ only depends on the dimension of the coordinate
$\fvec{\alpha}$. Additionally, there is a simple link, involving Hermite
polynomials, between the successive derivatives of a Gaussian shape and its
multiplication by polynomials. For instance, we have for the two first
derivatives in $\fvec{\alpha}$
\begin{align}
\label{eq:goa_overlap_derivatives}
  \frac{\partial\overlap^{1/2}}{\partial \fvec{\alpha}_k} &=
    -2 (\fvec{\alpha}_k - \fvec{\alpha'}_k)
    \overlap^{1/2}, \\
  \frac{\partial^{2}\overlap^{1/2}}{\partial \fvec{\alpha}_k \partial \fvec{\alpha}_l}  &=
    [ -2\delta_{kl}
     + 4(\fvec{\alpha}_k - \fvec{\alpha}'_k)
        (\fvec{\alpha}_l - \fvec{\alpha}'_l)]
    \overlap^{1/2}.
\end{align}

In the following, we build a local collective Hamiltonian. After the change of
variable~\eqref{eq:goa_change_var}, the Hamiltonian kernel between two
arbitrary GHW states reads
\begin{equation}\label{eq:goa_hkernel}
  \braket[\hat{H}]{\Psi}{\Phi} =
    \int_{\fvec{\alpha} \fvec{\alpha}' \fvec{\xi}}
      f^{\star}_{\Psi}(\fvec{\alpha}) \overlap^{1/2}(\fvec{\alpha}, \fvec{\xi})
    h(\fvec{\alpha},\fvec{\alpha}')
    \overlap^{1/2}(\fvec{\xi} , \fvec{\alpha}') 
    f_{\Phi}(\fvec{\alpha}')
    \diff{\fvec{\alpha}}\diff{\fvec{\alpha}'}\diff{\fvec{\xi}}.
\end{equation}
By assuming that the reduced Hamiltonian is a second-degree polynomial, we can
write down for any point $\fvec{\xi}$
\begin{multline}
  h(\fvec{\alpha},\fvec{\alpha}') =
      h(\fvec{\xi}, \fvec{\xi})
    + h_{\fvec{\alpha}} (\fvec{\alpha} - \fvec{\xi}) 
      + h_{\fvec{\alpha}'} (\fvec{\alpha}' - \fvec{\xi}) \\
    + \frac{1}{2}
      \big[ 
          h_{\fvec{\alpha}\fvec{\alpha}} (\fvec{\alpha} - \fvec{\xi})^{2} 
        + 2 h_{\fvec{\alpha \alpha}'} (\fvec{\alpha} - \fvec{\xi})(\fvec{\alpha}' - \fvec{\xi})
        + h_{\fvec{\alpha'\alpha'}} (\fvec{\alpha}'-\fvec{\xi})^{2}
      \big],
\end{multline}
where $h_{\fvec{\alpha}}$ is a shorthand notation for the vector of the first
derivatives of the reduced Hamiltonian estimated at $\fvec{\xi}$
\begin{equation}
  h_{\fvec{\alpha}} \equiv
  \left(
    \left.
      \frac{\partial h(\fvec{\alpha},\fvec{\alpha}')}{\partial \fvec{\alpha}_{1}}
    \right|_{\fvec{\alpha}=\fvec{\alpha}'=\fvec{\xi}},
    \dots,
    \left.
      \frac{\partial h(\fvec{\alpha},\fvec{\alpha}')}{\partial \fvec{\alpha}_{\Qdim}}
    \right|_{\fvec{\alpha}=\fvec{\alpha}'=\fvec{\xi}}
  \right).
\end{equation}
Similarly $h_{\fvec{\alpha}\fvec{\alpha}}$, $h_{\fvec{\alpha}\fvec{\alpha}'}$,
and $h_{\fvec{\alpha}'\fvec{\alpha}'}$ are the tensors of second derivatives
with respect to the collective coordinates and evaluated at point $\fvec{\xi}$.
The idea is then to inject this local development into
equation~\eqref{eq:goa_hkernel}. Using the
relation~\eqref{eq:goa_overlap_derivatives}, we express the reduced
kernel as a local operator containing derivatives acting on the right-hand side
$\nker^{1/2}$. Finally, after rearranging all the
terms and performing some integrations by parts, we obtain the expected result
\begin{equation}
  \braket[\hat{H}]{\Psi}{\Phi} =
    \int_{\fvec{\alpha}\fvec{\alpha}'}
      g^{\star}_{\Psi}(\fvec{\alpha})
        \col{H}(\fvec{\alpha}) \delta(\fvec{\alpha} - \fvec{\alpha}')
      g_{\Phi}(\fvec{\alpha}')
      \diff{\fvec{\alpha}} \diff{\fvec{\alpha}'}.
\end{equation}
The identification of this expression with~\eqref{eq:gcm:opmapping} shows that
the collective Hamiltonian is local. It reduces to a standard
kinetic-plus-potential Hamiltonian acting on the collective wave function
\begin{equation}
  \col{H}(\fvec{\alpha}) =
     -\frac{\hbar^2}{2} \nabla_{\fvec{\alpha}} B(\fvec{\alpha}) \nabla_{\fvec{\alpha}}
    + V(\fvec{\alpha}).
\end{equation}
The potential and inertia matrices in this coordinate representation are
\footnote{
  Note that some higher-order correction terms in the potential are neglected
  here (see Ref.~\cite{reinhardGeneratorcoordinate1987} for more details).
}
\begin{equation}\label{eq:all_coll}
  \begin{array}{l}
    \displaystyle
    V(\fvec{\alpha}) =
        h(\fvec{\alpha},\fvec{\alpha}) 
      - \frac{1}{2} \operatorname{Tr}(h_{\fvec{\alpha} \fvec{\alpha}'}) 
\medskip\\
    \displaystyle
    B(\fvec{\alpha}) =
        \frac{1}{2\hbar^2} ( h_{\fvec{\alpha}\fvec{\alpha}'}
      - h_{\fvec{\alpha} \fvec{\alpha}}).
  \end{array}
\end{equation}
Injecting this expression of the collective Hamiltonian
into~\eqref{eq:tdgcm_coll} and solving the resulting equation gives the
time-evolution of the unknown function $g(\fvec{\alpha})$. The
ultimate step is to transform back this equation of motion to another one
acting on the original set of coordinates $\fvec{q}$. Doing so, we get the same
equation with a transformed local collective Hamiltonian
\begin{equation}
\label{eq:goa_hcoll}
  \col{H}(\fvec{q}) =
   -\frac{\hbar ^2}{2\sqrt{\gamma(\fvec{q})}}
      \nabla_{\fvec{q}}\left[\sqrt{\gamma(\fvec{q})}B(\fvec{q})\right]\nabla_{\fvec{q}}
  + V(\fvec{q}).
\end{equation}
The new collective Hamiltonian involves a metric $\gamma(\fvec{q})$ defined by
\begin{equation}
\label{eq:goa_metric}
  \gamma(\fvec{q})=\operatorname{det}\left(G(\fvec{q}) \right).
\end{equation}
The inertia tensor takes the more involved form
\begin{equation}\label{eq:goa_inertia}
  B(\fvec{q}) =
    \frac{1}{2\hbar^2} G^{-1}(\fvec{q})\Big[
      h_{\fvec{q}\fvec{q}'} - h_{\fvec{q}\fvec{q}} 
    + \sum_n \Gamma^n(\fvec{q}) h_{\fvec{q}_n} 
    \Big]
    G^{-1}(\fvec{q}).
\end{equation}
The notation $\Gamma^n(\fvec{q})$ stands for the Christoffel symbol. It is a
matrix related to $G(\fvec{q})$ through the relation
\begin{equation}
  \Gamma^n_{kl}(\fvec{q}) =
    \frac{1}{2} \sum_i G^{-1}_{ni}\left(
      \frac{\partial G_{ki}}{\partial \fvec{q}_l}
    + \frac{\partial G_{il}}{\partial \fvec{q}_k}
    - \frac{\partial G_{lk}}{\partial \fvec{q}_i}
    \right).
\end{equation}
Finally, the potential becomes in this set of coordinate
\begin{equation}\label{eq:goa_potential}
  V(\fvec{q}) = 
    h(\fvec{q},\fvec{q})
  - \frac{1}{2} \operatorname{Tr}\left(
      G^{-1}(\fvec{q}) h_{\fvec{q} \fvec{q}'}
    \right).
\end{equation}
The first term is the HFB energy of the generator state $\kHFB{\fvec{q}}$. The
second term is a zero-point correction that contains second derivatives of the
reduced Hamiltonian. With some additional work, it is possible to express this
zero-point correction $\epsilon_{\text{ZPE}}$ in a slightly more practical form
that involves the inertia tensor and second derivatives of the energy
$h(\fvec{q},\fvec{q})$ only
\begin{equation}\label{eq:goa_zpe}
  \epsilon_{\text{ZPE}}(\fvec{q}) =
    - \frac{\hbar^2}{2}\operatorname{Tr}(B G)
    - \frac{1}{8}
        \operatorname{Tr}\left(
          G^{-1} \frac{\partial^2 h(\fvec{q},\fvec{q})}{\partial \fvec{q}^{2}}
        \right)
    + \frac{1}{8}\operatorname{Tr}\left(
        G^{-1} \sum_n \Gamma^n \frac{\partial h(\fvec{q},\fvec{q})}{\partial a_n}
      \right).
\end{equation}
The equation of evolution~\eqref{eq:tdgcm_coll} along with the expression
of the collective Hamiltonian~\eqref{eq:goa_hcoll} and its
components~\eqref{eq:goa_metric}, \eqref{eq:goa_potential} and
\eqref{eq:goa_inertia} define the dynamics of the system in the TDGCM+GOA
framework.

\subsubsection{Inertia and metric}
\label{sec:goa_cst}

The inertia tensor and the metric are quantities that depend on the derivatives
of the generator states and the reduced Hamiltonian. One possibility could be
to determine these derivatives numerically, for instance, with a finite
difference method. In the standard situation where the generator states are
constrained HFB solutions, one can find an analytical expression of the inertia
and the metric. We recall here this result at any point $\fvec{q}$
\begin{align}
\label{eq:goa_gb_moments}
  G &=
    \frac{1}{2} [\Mvec^{(1)}]^{-1}
                 \Mvec^{(2)}
                [\Mvec^{(1)}]^{-1}. \\
  B &=
           \Mvec^{(1)}
          [\Mvec^{(2)}]^{-1}
    \tilde{\Mvec}^{(1)} 
          [\Mvec^{(2)}]^{-1}
           \Mvec^{(1)}.
\end{align}
The moments $\Mvec^{(K)}$ and $\tilde{\Mvec}^{(K)}$ involve the QRPA matrix
$\mathcal{M}$ of the state $\kHFB{\fvec{q}}$ and are defined in
App.~\ref{app:goa_moments}. For the complete derivation of these results, we
refer the reader to~\cite{schunckMicroscopic2016} and references therein.
Note that this result neglects the term involving the Christoffel symbol in
the inertia. The argument for this approximation relies on the slow variation
of the metric according to the collective coordinates. We are not aware of the
systematic verification of the validity of this assumption in applications.

In all TDGCM+GOA practical applications, the so-called perturbative cranking
approximation is used to avoid a costly inversion of the QRPA matrix required
to compute the metric and inertia. It consists in approximating the QRPA matrix
by a diagonal part only, in the quasiparticle basis that diagonalizes the
generalized density matrix of $\kHFB{\fvec{q}}$. This gives a simple and well
known form for the moments $\Mvec^{(K)}$
\begin{equation}
  M_{ij}^{(K)} = \tilde{M}_{ij}^{(K)} =
    \mathfrak{Re} 
      \sum_{\mu\nu}
        \frac{\braket[\hat{Q}_{i}]{\mu\nu}{ \HFB(\fvec{q})}
              \braket[\hat{Q}_{j}]{ \HFB(\fvec{q}) }{\mu\nu}}{(E_{\mu}+E_{\nu})^{K}},
\end{equation}
where $\ket{\mu\nu}$ is a two quasiparticles excitation built on top of the
generator state, and $E_{\mu}$ and $E_{\nu}$ are the corresponding
quasiparticle energies.

The GCM+GOA framework unambiguously defines the metric and inertia as functions
of the successive derivatives of the generator states and reduced Hamiltonian.
However, it is known that this inertia and its approximate perturbative cranking
estimation is too low to describe several situations correctly. One example
is the case of a translation motion~\cite{reinhardGeneratorcoordinate1987}.
Several studies compare the GOA inertia with inertia provided by other theories
yielding an equivalent collective equation of motion, such as quantized
ATDHFB~\cite{holzwarthFour1973,gozdzMass1985,baranQuadrupole2011}. In
Ref.~\cite{reinhardGeneration1978}, the authors extend the TDGCM+GOA framework
by introducing conjugate coordinates that bring time odd components into the
generator states. In particular, they show that the resulting collective
Hamiltonian takes the same form as Eq.~\eqref{eq:goa_hcoll} but where the
ATDHFB inertia replaces the GOA inertia. This justifies the common practice of
using the ATDHFB inertia when solving the collective equation of motion.

\subsection{Applications in nuclear reactions}

\subsubsection{Low energy ion collisions}

\begin{figure}[!ht]
\begin{center}
  \includegraphics[width=0.5\textwidth]{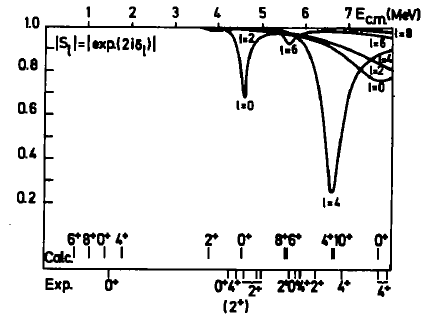}
  \caption{
    Reflection coefficients as a function of the center of mass energy (in MeV)
    of a $^{12}$C+$^{12}$C head-on collision. The position of the resonances
    estimated from a GCM+GOA framework are compared to a set of experimental
    data. Figure taken from Ref.~\cite{bergerSelf1980}.
  }
  \label{fig:berger_c12c12}
\end{center}
\end{figure}

The force of TDGCM+GOA is its versatility in the choice of collective
coordinates and its ability to treat in the same framework the nucleons DoFs as
well as more collective DoFs. It seems an appropriate way to tackle the
dynamics of low energy ion collisions where the principal degree of freedom is
the relative distance between the two reaction partners and where the collision
affects the internal organization of the nucleons. It is possible to build a
family of generator states along this line, describing the two reaction
partners and parametrizing them by their relative distance. Several papers
followed this idea during the 1980s. In particular, J.-F.~Berger and
D.~Gogny~\cite{bergerSelf1980} treated the frontal collision of
$^{12}$C+$^{12}$C within a GCM+GOA approach. This kind of study focuses on the
determination of the cross-section resonances for some specific output channels
of the reaction. Figure~\ref{fig:berger_c12c12} shows a typical result where
the resulting positions of the resonances are compared to available
experimental data. The predictions give a rough estimation of the position of
the $0^+$ resonances, but they mostly fail to reproduce the presence of other
resonances and their energy spacing. Many lacunae of the theory could explain
such discrepancy, including the rough treatment of angular momentum, the
breaking of some symmetries, or the mostly adiabatic characteristic of the GCM
built on constrained HFB solutions.  
Other similar studies have been performed on the base of the GCM (without the
GOA) and have made the connection to the resonating group method. D.~Baye and
Y.~Salmon looked at the $^{16}$O+$^{40}$Ca back angles
scattering~\cite{bayeGenerator1979} along with the work of Friedrich
\textit{et al.}~\cite{friedrichElastic1976}. Also, Goeke \textit{et al.}
studied the $^{16}$O+$^{16}$O collision in the framework of the quantized
adiabatic time-dependent Hartree-Fock approach which yields a collective
equation of motion identical to the one of
TDGCM+GOA~\cite{goekeThreedimensional1983}.

After this series of applications, treating collisions with the TDGCM+GOA
framework was progressively abandoned to the profit of other methods such as
the time-dependent Hartree-Fock plus pairing~\cite{sekizawaTDHF2019}. One
difficulty that could explain this transition is the numerical cost required to
build the generator states at the self-consistent mean-field level (note that
this cost is nowadays completely acceptable). Beyond this, deeper problems
raised, for instance, by the conservation of the total angular momentum of the
collision or the generation of a continuous manifold of generator states appear
with this method. Overall, the resulting cross-sections give only rough and
qualitative estimations of the experimental data. The position of resonances,
as well as the absolute value of cross-sections, are both observables that are
very challenging to predict due to their extreme sensitivity to the kinematics
of the reaction as well as the internal structure of the nuclei.

\subsubsection{Fission dynamics}

The prediction of the fission fragments characteristics from a dynamical
description is a domain where the TDGCM+GOA performs successfully. Fission
involves heavy nuclei and begins with large collective motions that are mostly
adiabatic. These two factors make the TDGCM+GOA framework built on constrained
HFB solutions a suitable candidate. Moreover, the important width of the
measured fission yields is the fingerprint of large quantum fluctuations of the
one-body density of the compound system. Handling these fluctuations is
precisely the purpose of the GCM.

The quest to predict fission yields from a dynamical TDGCM+GOA calculation
began in the 1980s with the work of J.-F. Berger \textit{et al.} exploring the
rupture of the neck between prefragments in terms of different collective
coordinates~\cite{bergerMicroscopic1984,bergerTimedependent1991}. The first
calculation of the mass distribution of fission fragments was later on
performed among the same group for $^{238}$U~\cite{goutteMicroscopic2005}. The
authors have described the fissioning system's dynamics using the two
collective coordinates: $q_{20}$ and $q_{30}$ associated with the quadrupole
and octupole moments of the compound nucleus. The Ref.\cite{younesFragment2012}
reports the same technique applied to a few other actinides with a qualitative
reproduction of the experimental values. W.~Younes and D.~Gogny further
proposed an alternative set of collective variables
in~\cite{younesCollective2012}. Still, an impediment to this approach was its
numerical cost, from the determination of the generator states (up to $40000$
states in a 2-dimensional description) to the time integration of the
collective Schr\"odinger equation. The development of new tools based on
state-of-the-art numerical methods enables today's continuation of this work.
For instance, the code FELIX~\cite{regnierFelix12016,regnierFelix22018} solves
the collective GOA dynamics efficiently based on a spectral element method.
Also, the use of Bayesian processes to determine the best-suited
parameters of a harmonic oscillator basis induced a significant speedup of some
Hartree-Fock-Bogoliubov solvers.

\begin{figure}[ht!]
\begin{tabular}{cc}
 \includegraphics[width=0.45\textwidth]{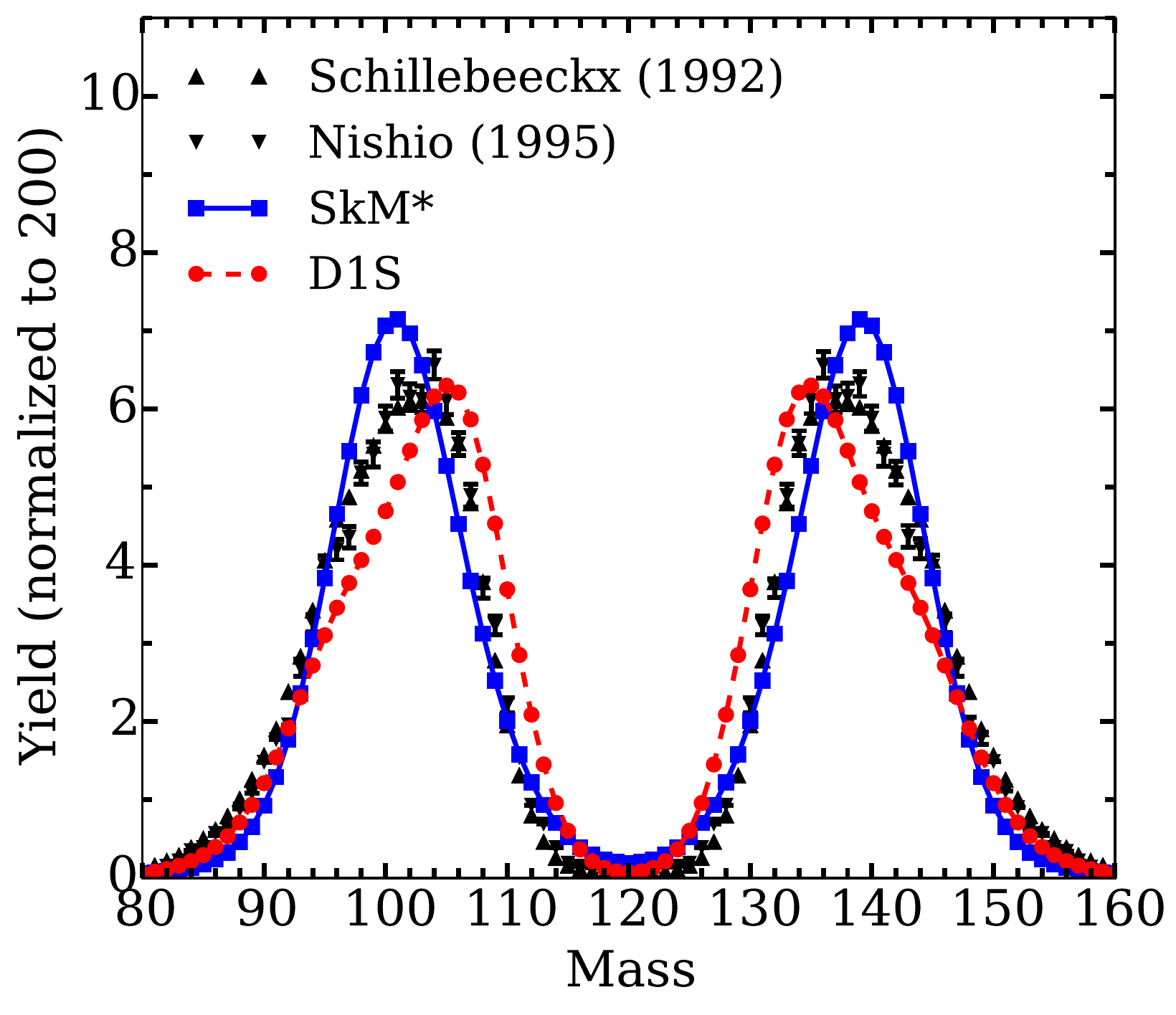}
 \includegraphics[width=0.51\textwidth]{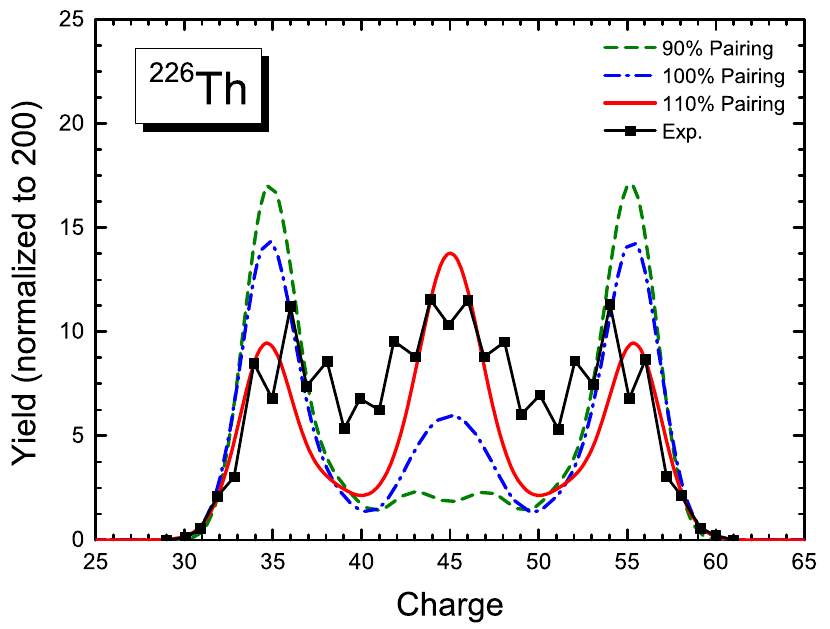}
\end{tabular}
 \caption{Left panel:
 Fragment mass distribution for the low energy neutron induced fission $^{239}$Pu(n,f).
 Two TDGCM results obtained with the Gogny D1S and Skyrme SkM$^*$ effective interactions are compared to two sets of experimental data.
 Reprinted figure with permission from Ref.~\cite{regnierFission2016}.
 Copyright 2016 by the American Physical Society.
 Right panel:
 Fragment charge distribution obtained for a low energy fission of $^{226}$Th.
 The TDGCM+GOA results based on the relativistic mean field PC-PK1 with different pairing strengths
 are compared to experimental data (black line with points).
 Reprinted figure with permission from Ref.~\cite{taoMicroscopic2017}. 
 Copyright 2017 by the American Physical Society.}
 \label{fig:yA}
\end{figure}

\begin{figure}[!ht]
\begin{tabular}{cc}
  \includegraphics[width=0.49\textwidth]{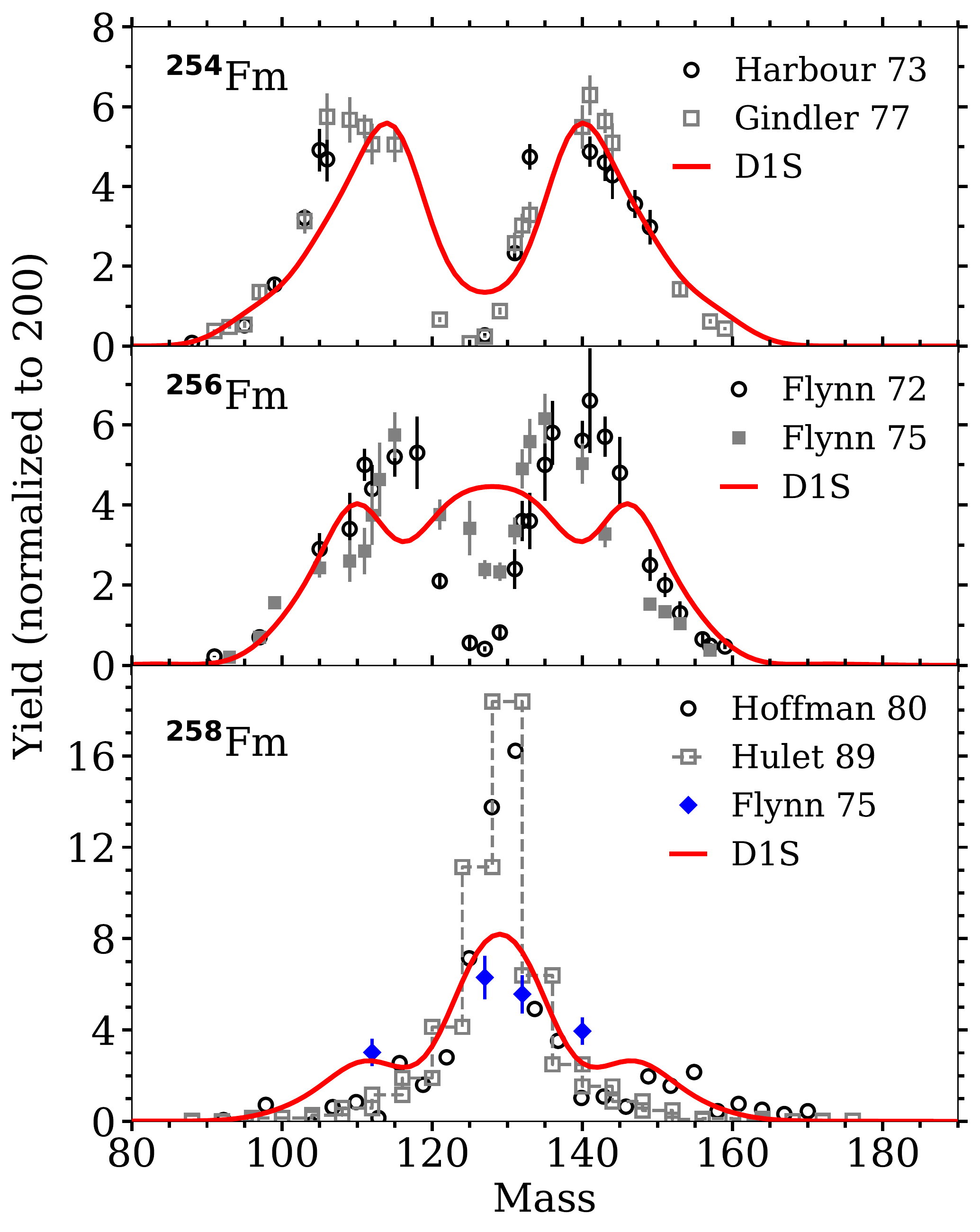} &
  \includegraphics[width=0.45\textwidth]{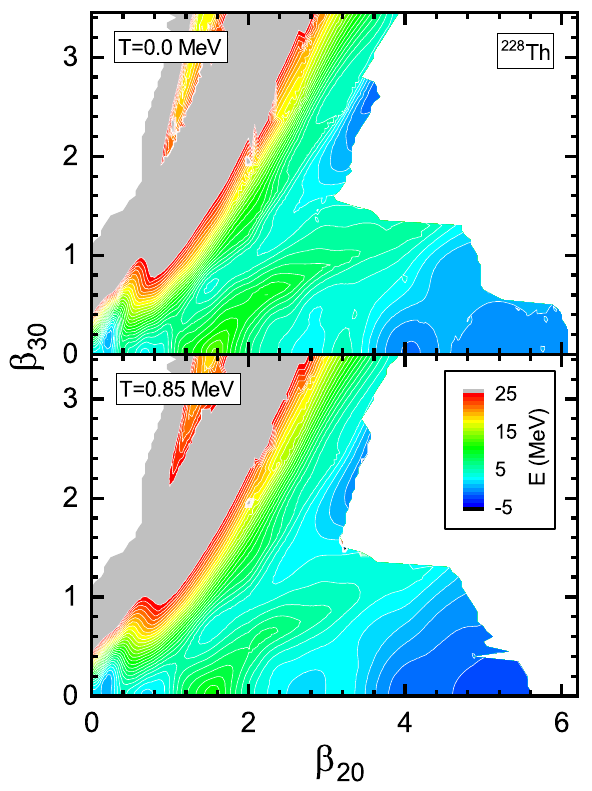}
\end{tabular}
  \caption{
    Left panel:
    Primary fragment mass yields of Fermium isotopes obtained with the Gogny
    D1S effective interaction and compared with various experimental data sets
    after neutron evaporation. The open symbols stand for experimental data
    associated with spontaneous fission, whereas full symbols are related to
    thermal neutron-induced fission.
    Reprinted figure with permission from Ref.~\cite{regnierAsymmetric2019}. 
    Copyright 2019 by the American Physical Society.
    Right panel:
    Effect of temperature on the free energy surface of $^{228}$Th in the
    plane of deformation $(\beta_{20},\beta_{30})$.
    Reprinted figure with permission from Ref.~\cite{zhaoTimedependent2019}. 
    Copyright 2019 by the American Physical Society.
    }
  \label{fig:theovsexp}
\end{figure}

In the last couple of years, we have seen a fast increase in the number of
fission studies relying on this technique. All papers focused on the actinide
region emphasize similar results. In this region, the potential energy
landscape presents mostly one asymmetric fission valley. The exact topology of
this surface reflects the internal organization of the nucleons that would
correspond to the shell effects in a microscopic-macroscopic picture. By
starting from a collective state localized in the low deformation first
potential well, the dynamics mostly populates the configurations of the
asymmetric channel. The left panel of Fig.~\ref{fig:yA} shows the resulting
yields obtained on an experimentally well-known nucleus, namely $^{240}$Pu.
The TDGCM+GOA captures within a few mass units the position of the asymmetric
peaks that are, in fact, mostly determined by the position of the asymmetric
valley in the collective space. A similar quality of results has been obtained
with the same method for other actinides, such as $^{236}$U or $^{252}$Cf.
Finally, this framework seems to be able to describe the transitions between
symmetric and asymmetric fissions that are measured outside of the actinide
region. The left panel of Fig.~\ref{fig:theovsexp} shows the prediction versus
experiment comparison of such a transition in the neutron-rich Fermium
isotopes~\cite{regnierAsymmetric2019}. In this chain of isotopes, the addition
of a few neutrons to $^{254}$Fm changes the dominant fragmentation mode
completely. This can be interpreted as different shell effects occurring
because of the new neutrons, that change the potential energy of the
intermediate configurations leading to fission. This perturbation favors the
population of the symmetric mode for $^{258}$Fm.

Several ingredients of the TDGCM+GOA framework for fission are still not
adequately controlled and bring significant uncertainties on its predictions.
In Ref.~\cite{zdebFission2017}, A.~Zdeb \textit{et al.} investigated in detail
the impact of the choice of the initial state of the dynamics on the fission
observables. They showed in particular that the global features of the fission
yields (mostly the position and width of the peaks) are quite resilient to
changes in the energy or the parity of the initial state. Furthermore,
Tao \textit{et al.} computed the fission yields from a relativistic mean-field
approach~\cite{taoMicroscopic2017} and looked at the sensitivity of the results
to the pairing strength. The right panel of Fig.~\ref{fig:yA} gives a clue of
their results, showing the variation of the charge yields induced by a 10\%
variation of their nominal pairing strength in the case of the multimodal
fission of $^{226}$Th. For this nucleus, we see that the pairing strength is an
essential factor that drives the ratio between the yields of the symmetric and
asymmetric modes. Finally, the same team explored the inclusion of temperature
into the generator states as a way to better account for the diabatic aspects
of the dynamics~\cite{zhaoMicroscopic2019,zhaoTimedependent2019}. The
Fig.~\ref{fig:theovsexp} (right panel) shows that warming up the generator
states changes slightly the topology of the potential energy surface.
Increasing the temperature generally tends to smear out the shell effects and
the structures in the potential energy surface. In the case of $^{226}$Th, it
favors the symmetric fission and reduces the height of the asymmetric peaks of
the mass yields by a factor $\simeq$1.4.
Other components or approximations of the TDGCM+GOA, such as the perturbative
cranking approximation for the collective inertia, may also bring their source
of bias and uncertainty on the prediction.

\subsection{Main limitations}
\label{sec:goa_limitations}

Despite its success in determining the fission fragment distribution, the
TDGCM+GOA framework suffers from several shortcomings.

First, on the same ground as the exact TDGCM, its derivation relies on the
knowledge of a many-body Hamiltonian. However, in all practical applications,
it is used with an energy density functional (cf.
Sec.~\ref{sec:multireference}). Indeed, the GOA method does not require an
explicit calculation of the off-diagonal elements of the energy kernel
responsible for divergent behavior in GCM. However, the GOA's formal
construction still depends on the existence and sound mathematical definition
of these matrix elements to be a valid framework. In that sense, the GOA
suffers from the same flaws as the exact TDGCM concerning the use of energy
density functionals.

A second issue comes from the requirement that the function $\fvec{q}
\rightarrow \kHFB{\fvec{q}}$ is continuous and twice differentiable. The latter
is a necessary condition to develop the formalism and, in particular, to
compute the GOA metric and inertia. However, the standard construction of the
family of generator states from constrained HFB solutions does not guaranty
this property~\cite{dubrayNumerical2012}. Different studies highlight
discontinuities of this function in the treatment of fission, similar to the
one visible in Fig.~\ref{fig:GCM:240PuN}. In the common $(q_{20},q_{30})$ space
of collective coordinates, a line of discontinuity is present in the vicinity
of scission configurations. This feature limits the domain of validity of the
collective dynamics and ultimately prevents the determination of the fission
fragments characteristics after their complete separation.

Finally, we have seen that most of the current applications of TDGCM+GOA rely
on constrained HFB solutions for the generator states. Certain diabatic aspects
of the nuclear dynamics are then difficult to grasp with the corresponding GHW
many-body wave function. This is the case of the dissipation as well as the
viscosity of the shape dynamics predicted with Langevin
methods~\cite{usangCorrelated2019,miyamotoOrigin2019} or time-dependent
Hartree-Fock-Bogoliubov calculations. Past and ongoing studies to improve the
description of these effects include efforts to quantize the Langevin
equation~\cite{parisiPerturbation1981,damgaardStochastic1984}, to couple the
Langevin dynamics with the GCM~\cite{sadhukhanMicroscopic2016} or to couple
TDHFB trajectories with TDGCM~\cite{bulgacFission2019}. Other techniques, such
as the SCIM and TDGCM based on time-dependent generator states, are also
promising avenues that we discuss in this review.

\section{Schr\"odinger Collective-Intrinsic Model (SCIM)}
\label{sec:scim}

Intrinsic degrees of freedom are often neglected in the microscopic modeling of
the dynamics of reactions. However, including intrinsic degrees of
freedom in a static GCM framework has already been performed, for instance, in
Refs.~\cite{mutherSingle1977,chenTriaxial2017}. These studies show that taking
into account two-quasiparticle excitations significantly improves the
prediction of high spin levels, such as the $6^+$ states in medium mass
isotopes as well as the prediction for $\beta$ excitation bands and its
transition probabilities to other rotational bands in heavier systems. On
another topic, the TDHFB/SLDA methods~\cite{bulgacInduced2016,
bulgacUnitary2019} and semi-classical approaches to the description of
fission~\cite{sierkLangevin2017,wadaMulti1992} clue that dissipation (and
therefore intrinsic degrees of freedom) are necessary to describe the fission
fragments properties correctly. Therefore, a few collective degrees of freedom
are not enough to adequately model such a reaction. Several paths can be taken
to overcome this limitation without resorting to the determination of an exact
solution of the GHW equation~\eqref{eq:gcm:stdHW}. A strategy in development
consists in using the TDGCM+GOA with finite-temperature inertia tensors and
collective potential. However, the inclusion of statistical mechanics on top of
the TDGCM framework still lacks a solid formalization. The idea of the
Schr\"odinger Collective-Intrinsic Model (SCIM)~\cite{bernardTaking2011,
bernardMicroscopic2011,younesMicroscopic2019} is to derive a local
Schrodinger-like equation from a generalization of the GHW
ansatz~\eqref{eq:gcm:GHW} that contains individual quasiparticle degrees of
freedom. The transformation to a local equation relies on the symmetric moment
expansion method~\cite{holzwarthConnection1972,kermanQuantum1974}. The full
SCIM formalism can be found in~\cite{bernardTaking2011,bernardMicroscopic2011,
younesMicroscopic2019} in the stationary case. However, we would like to
present here a derivation of the time-dependent SCIM equations consistent with
the ones given for the TDGCM and the TDGCM+GOA equations.

\subsection{Main assumptions}

\begin{figure}[ht]
  \includegraphics[width=0.49\linewidth]{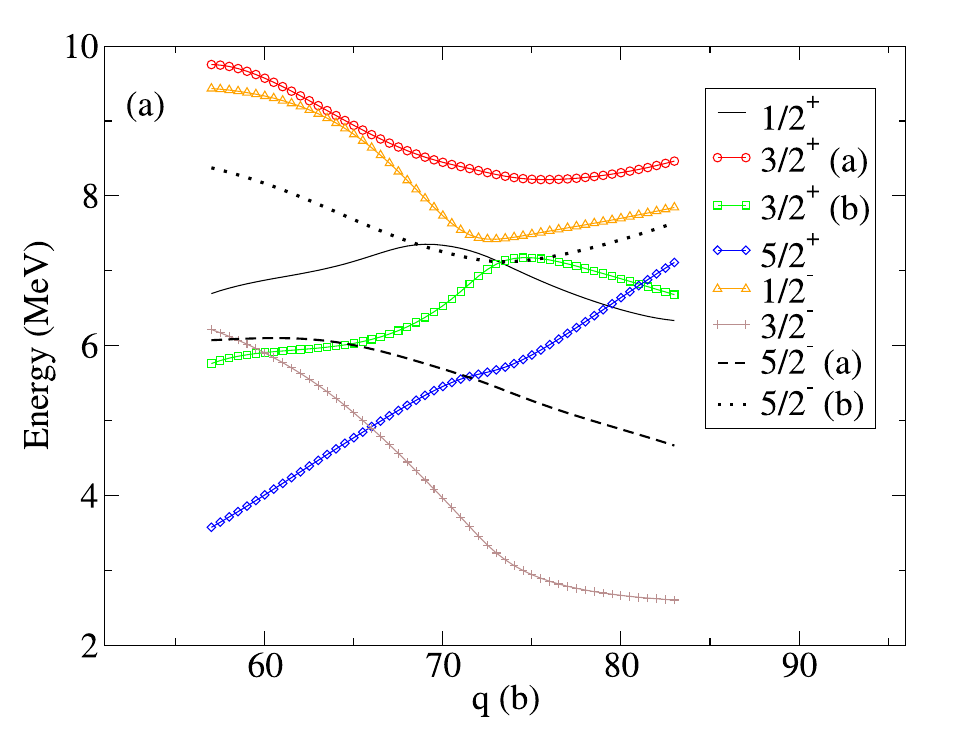}
  \includegraphics[width=0.49\linewidth]{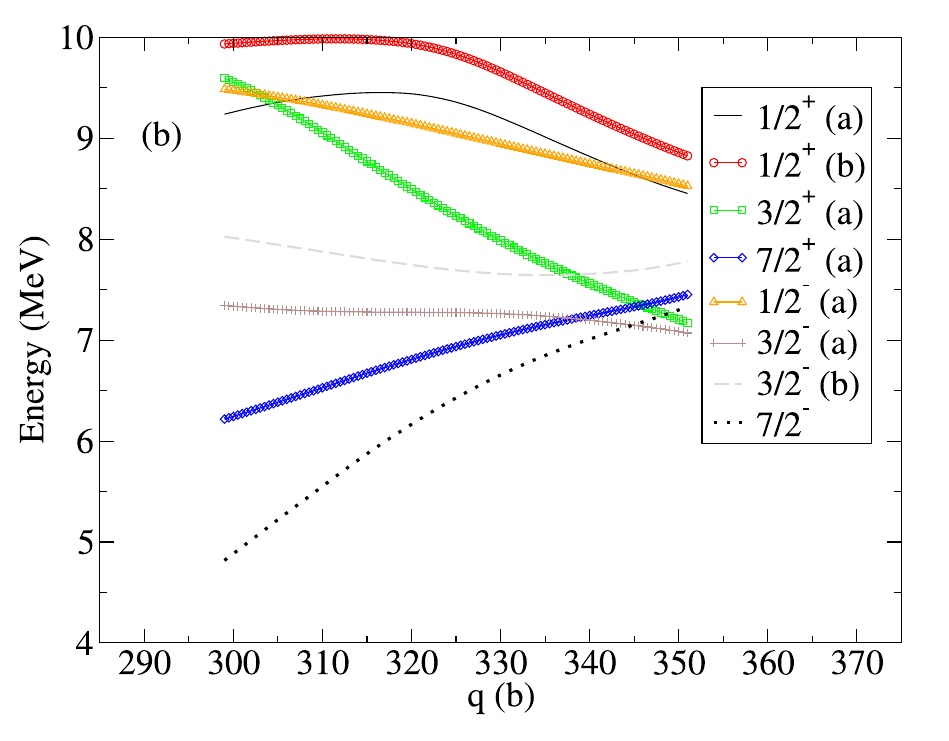}
  \caption{(Color online)
    Excitation energies as a function of the quadrupole moment constraint $q$
    and associated with 2-qp excitations of HFB states. The system under study
    is  \textsuperscript{236}U around $\qavg{\hat{Q}_{20}}=70$ b (figure a) and
    $\qavg{\hat{Q}_{20}}=325$ b (figure b). The figures are taken from
    Ref.~\cite{bernardTaking2011}.
  }
  \label{fig:non-adiab-pes}
\end{figure}

The SCIM involves four main assumptions. The first one is the expression of the
state $\kTDGCM{t}$ that describes the evolution of the many-body wavefunction
associated with the reaction. This expression is assumed to be a generalization
of the GHW ansatz
\begin{equation}\label{eq:SCIM:GHWgen}
  \kTDGCM{t} =
    \sum_k\int\diff{\fvec{q}}
      \kqpHFB{\fvec{q}}{k}
      f_k(\fvec{q}, t).
\end{equation}
In~\cite{bernardMicroscopic2011}, the authors consider a family of generator
states associated with one collective coordinate $q$ defined as the
quadrupole moment of the system. The index $k$ iterates over the labels of the
sheets of collective space which correspond, in this case, to two quasiparticle
excitations. Fig.~\ref{fig:non-adiab-pes} shows the evolution of the
excitation energies of the non-adiabatic points of the potential energy surface
of $^{236}$U.
Note that the scope of expression~\eqref{eq:SCIM:GHWgen} is broader than the
only explicit inclusion of intrinsic DoF in the formalism. For example, it is
used for $K$-mixing in the context of stationary angular-momentum-projected GCM
on a triaxial configuration basis. In this case, the index $k$ iterates over
the values of $K$. The second assumption is the analyticity of the weight
function $f$ of the GCM ansatz~\eqref{eq:SCIM:GHWgen} that allows the
symmetrization of the GHW equations. The third assumption is the vanishing of
the weight function and its derivatives at the boundaries of the integration
domain. An implicit corollary of this property is the continuity of the
functions $\fvec{q} \rightarrow \kqpHFB{\fvec{q}}{k}$. It turns out that this
assumption is in practice not verified for a broad range of applications, for
example, in the actinide region, as emphasized in
Sec.~\ref{sec:goa_limitations}. These three assumptions lead to the symmetrized
GHW equation
\begin{equation}\label{eq:SCIM:HWsym}
  \sum_k\int\diff{\fvec{s}}\;
      e^{i\fvec{s}\fvec{\opRep{P}}/2}
      \left[
        \hker[\fvec{s}]_{lk}(\fvec{q})
      - i\nker[\fvec{s}]_{lk}(\fvec{q})\frac{\diff{}}{\diff{t}}
      \right]
      e^{i\fvec{s}\fvec{\opRep{P}}/2}
    f_k(\fvec{q}, t)
      = 0,
\end{equation}
where the following notations are introduced
\begin{align}
  \label{eq:SCIM:hker}
  \hker[\fvec{s}]_{lk}(\fvec{q})
    &= \hker_{lk}(\fvec{q}+\fvec{s}/2, \fvec{q}-\fvec{s}/2) \\
  \label{eq:SCIM:Nker}
  \nker[\fvec{s}]_{lk}(\fvec{q})
    &= \nker_{lk}(\fvec{q}+\fvec{s}/2, \fvec{q}-\fvec{s}/2),
\end{align}
and where the Hermitian operator
\begin{equation}
  \fvec{\opRep{P}} = i\frac{\partial}{\partial \fvec{q}}
\end{equation}
corresponds to the conjugate moment associated with the collective variables.
The symmetrized GHW equation can be written in a more compact operator format
as
\begin{equation}\label{eq:SCIM:HWsym2}
  \int\diff{\fvec{s}}\;
      e^{i\fvec{s}\fvec{\opRep{P}}/2}
      \left[
        \hker[\fvec{s}]
      - i\nker[\fvec{s}]\frac{\diff{}}{\diff{t}}
      \right]
      e^{i\fvec{s}\fvec{\opRep{P}}/2}
    \vecRep{f}(t)
      = 0,
\end{equation}
where $\vecRep{f}(t)$ denotes the function $\fvec{q} \mapsto f_k(\fvec{q},t)$.
The fourth and last assumption of the SCIM is the validity of the truncation of the
symmetric moment expansion (SME) of the norm and Hamiltonian kernels
of~\eqref{eq:SCIM:HWsym2} up to order two. It was, for instance, verified
numerically in the context of the study~\cite{bernardMicroscopic2011}. The SME
of $\kop=\nop,\hop$, in the case of one collective variable,
\begin{equation}\label{eq:SCIM:fullSME1D}
  \kop =
    \sum_{n}
      \frac{1}{n!}
      \SOPO{n}{\kop^{(n)}}{\opRep{P}},
\end{equation}
is obtained through the properties of the so-called Symmetric Ordered Product
of Operators (SOPO) $\SOPO{n}{\kop^{(n)}}{\opRep{P}}$ presented
in App.~\ref{app:SOPO} where $\kop^{(n)}$ is the moment of order $n$ of $\kop[s]$ in
the variable $s$
\begin{equation}\label{eq:SCIM:kmoments}
  \kop^{(n)} \equiv
    i^n\int\diff{s} s^n \kop[s].
\end{equation}
The properties of the SOPO used to obtain these expressions are listed in App.~\ref{app:SOPO}.

The expression~\eqref{eq:SCIM:fullSME1D} can be generalized to the case of
$\Qdim$ collective variables,
\begin{equation}\label{eq:SCIM:fullSMEND}
  \kop =
    \sum_{\fvec{n}}
      \frac{1}{\fvec{n}!}
      \SOPO{\fvec{n}}{\kop^{(\fvec{n})}}{\fvec{\opRep{P}}},
\end{equation}
where the index $\fvec{n}$ iterates over all the $\Qdim$-tuples of positive
integers and where we have introduced the following notations
\begin{align}
  \fvec{n}! &\equiv
    n_0!n_1!\cdots n_{\Qdim-1}! \\
  \SOPO{\fvec{n}}{\kop}{\fvec{\opRep{P}}} &\equiv
    \SOPO{n_{\Qdim-1}}{\cdots
      \SOPO{n_1}{\SOPO{n_0}{\kop}{\opRep{P}_0}}{\opRep{P}_1}
      \cdots}{\opRep{P}_{\Qdim-1}} \\
  \opRep{K}^{(\fvec{n})} &\equiv
    \int\diff{\fvec{s}} \left[\prod_k(is_k)^{n_k}\right]\opRep[\fvec{s}]{K}.
\end{align}

Their second-order approximation in their SME development is then given by
\begin{align}
  \nop &\approx
    \sum_{\bar{\fvec{n}}\leq 2}
      \frac{1}{\fvec{n}!}
      \SOPO{\fvec{n}}{\nop^{(\fvec{n})}}{\fvec{\opRep{P}}} \\
  \hop &\approx
    \sum_{\bar{\fvec{n}}\leq 2}
      \frac{1}{\fvec{n}!}
      \SOPO{\fvec{n}}{\hop^{(\fvec{n})}}{\fvec{\opRep{P}}},
\end{align}
where $\bar{\fvec{n}}$ is the sum of the elements of $\fvec{n}$. In
the one-dimensional case, the expressions reduce to
\begin{align}
  \nop &\approx
    \sum_{n=0}^2
      \frac{1}{n!}
      \SOPO{n}{\nop^{(n)}}{\opRep{P}}, \\
  \hop &\approx
    \sum_{n=0}^2
      \frac{1}{n!}
      \SOPO{n}{\hop^{(n)}}{\opRep{P}}.
\end{align}

\subsection{Schrodinger Collective-Intrinsic Equation}
\label{subsec:scim:SCIE}

The Schrödinger-like expression of the SCIM equations is given by
\begin{equation}\label{eq:SCIM:reducedSME}
  \left[
    \hop^{\rm CI}
  - i\frac{\diff{}}{\diff{t}}
  \right]
  \vecRep{g}(t) = 0,
\end{equation}
where $\vecRep{g}(t)$ is defined according to
\begin{equation}
  \vecRep{g}(t) = \nop^{1/2}\vecRep{f}(t)
\end{equation}
and normalized as
\begin{equation}
    \vecRep{g}^\dagger(t)\vecRep{g}(t)
  = \int \diff{\fvec{q}} g^{\star}(\fvec{q}, t)g(\fvec{q},t)
  = 1.
\end{equation}
The operator $\nop^{1/2}$ is the only positive-definite hermitian square-root
of $\nop^{1/2}$ and $\nop^{-1/2}$ is the inverse of the latter. Finally, using
the hermicity of $\nop^{-1/2}$, the collective-intrinsic Hamiltonian
$\hop^{\rm CI}$ has the expression
\begin{equation}
  \hop^{\rm CI} = \nop^{-1/2}\hop\nop^{-1/2}.
\end{equation}

An explicit form for $\hop^{\rm CI}(q)$ is given by
\begin{equation}\label{eq:SCIM:HamiltCI}
\opRep{H}^{\rm CI} =
      \frac{1}{2}\SOPO{2}{\opRep{B}}{\opRep{P}}
    + \SOPO{1}{\opRep{T}}{\opRep{P}}
    + \opRep{V},
\end{equation}
where the expressions of $\opRep{U}=\opRep{B}/2$, $\opRep{T}$ and $\opRep{V}$
are given in~\cite{bernardTaking2011,bernardMicroscopic2011,
younesMicroscopic2019}. By analogy with the TDGCM+GOA collective
Hamiltonian~\eqref{eq:goa_hcoll}, the first term of~\eqref{eq:SCIM:HamiltCI}
can be interpreted as a kinetic term and $\opRep{B}$ as the inertia tensor, related to
the mass tensor $\opRep{M}$ through the relation
\begin{equation}
  \opRep{B} = \opRep{M}^{-1}.
\end{equation}
Similarly, the third term of~\eqref{eq:SCIM:HamiltCI} is comparable to the
potential term of the TDGCM+GOA. However, the last term
\begin{equation}
  \SOPO{1}{\opRep{T}}{\opRep{P}} =
    \frac{1}{2}
    \left[
      \opRep{T}\frac{\partial}{\partial q}
    + \frac{\partial}{\partial q}\opRep{T}
    \right],
\end{equation}
contains first-order derivatives according to the collective variable, at the
opposite of the TDGCM+GOA. In the Langevin equations, such a term corresponds
to viscosity and arises in the SCIM from the coupling between intrinsic
and collective degrees of freedom.

\subsection{Choice of quasiparticle excitations}
\label{subsec:QPchoice}

In~\cite{bernardMicroscopic2011,bernardTaking2011,younesMicroscopic2019}, the
generator states consist in
\begin{itemize}
  \item constrained HFB states $\ket{\HFB_{k=0}(q=\qavg{\hat{Q}_{20}})}$ describing
    the compound system at different elongations,
  \item intrinsic excitations of these HFB states
    \begin{equation}
      \ket{\HFB_{k>0}(q)} =
        \hat{X}(q)_k\ket{\HFB_{0}(q)}.
    \end{equation}
\end{itemize}
Note that the specific expression of $\hat{X}(q)_k$ is never used in the
derivations of the Schrodinger-like equation, and it is only assumed that all
the states in the collective space are time-reversal to avoid complex-valued
overlaps. In practice, the intrinsic excitations taken into account in the
existing developments of SCIM are considering 2-qp excitations. The included
HFB states are breaking the rotational and particle number symmetries. In order
to avoid restoring these symmetries, the quasiparticle excitations are chosen
according to the following rules
\begin{enumerate}
  \item the operators $\hat{X}_k$ are two quasiparticles operators,
  \item all the states in the collective space have to be time-reversal invariant,
  \item the chosen excitations have to preserve ``as much as possible'' the
    number of particles and $K$, the projection of the total angular moment
    on the symmetry axis, \item and they must be associated with an excitation
    energy below 10~MeV. 
\end{enumerate}

The time-reversal condition limits the possible excitation operators to be
\begin{align}
  \hat{X}(q)_k  &=
      \alpha_{k}\cQPw[q]{k_1}\cQPw[q]{\bar{k}_2}
    - \cQPw[q]{\bar{k}_1}\cQPw[q]{k_2} \\
  \alpha_{k} &=
      \frac{1}{\sqrt{2}}
      \left(
        1+\delta_{k_1,k_2}
          \left(
            1 - \frac{1}{\sqrt{2}}
          \right)
      \right),
\end{align}
where $\cQPw[q]{k_1}$ is the creation operator of the quasiparticle $k_1$
associated with the HFB state
\begin{equation}
  \ket{\HFB_{0}(q)} =
    \prod_l \aQPw[q]{l}\ket{0}.
\end{equation}
Additionally, the selected quasiparticles in $\hat{X}(q)_k$ are assumed to
have the same projection on the total angular moment on the symmetry axis
$K_{k_1} = K_{k_2}$ so that the $K$ of the total system is unchanged. In case
the HFB states are obtained with preserved parity, the same condition on $\pi$
is added.

Couplings between collective and intrinsic excitations play a major role in many
reactions. For instance, it is known to play a crucial role in the distribution
of excitation energy between the nascent fragments produced by fission. The
TDGCM+GOA enables a microscopic description of nuclear reactions without
internal degrees of freedom, while Langevin-based methods allow the
semi-classical description of the reaction with the inclusion of thermal
effects. The SCIM leads to a local Schrodinger-like equation, much simpler to
solve than the exact, non-local, Griffin-Hill-Wheeler equation while being based
on fewer assumptions than the TDGCM+GOA or Langevin. The collective-intrinsic
Hamiltonian includes a viscosity term that is known to be relevant to the
description of nuclear reaction from Langevin's calculations. However, the
method still involves the full calculation of the norm and Hamiltonian kernels,
which is extremely time-consuming. Furthermore, the formalism is rather complex
compared to other methods such as the TDGCM+GOA. At present, this method did
not lead to any application beyond the works presented
in~\cite{bernardTaking2011,bernardMicroscopic2011,younesMicroscopic2019}, and
still needs to be tested thoroughly against experimental data.

\section{Quantum mixture of time-dependent states}
\label{sec:MCTDHF}
In its standard form, the TDGCM relies on the ansatz~\eqref{eq:gcm:GHW}
that expands the many-body wave function on a family of time-independent
generator states. The dynamics of the system is, therefore, entirely carried
out by the time evolution of the collective wave function $g(\fvec{q})$ driven
by Eq.~\ref{eq:tdgcm_coll}. Although successful in describing some nuclear
phenomena like collective vibrations, such an expansion suffers from two
significant drawbacks.

The first one resides in the large dimension of the ensemble of generator states
required to describe processes like nuclear reactions correctly. Despite the
efforts reported in Sec.~\ref{sec:goa} and~\ref{sec:scim} to reduce the
collective Hamiltonian to a local approximation, this high dimension quickly
becomes a hindrance to the numerical applications of TDGCM. An origin of this
difficulty is the fact that all the many-body configurations populated at any
time of the reaction must be represented in the set of generator states.
In many situations, this expansion is not optimal in the sense that most of the
associated weights are close to zero at a given time. To give an example, we
may consider the translation motion of a localized particle. While the
translated states at any positions are to be incorporated in a TDGCM
description of its motion, the collective wave function at a given time only
has a small spatial expansion. A natural idea is then to express the wave
function as a linear superposition of a few time-dependent states that follow
the expected particle's translation motion. It may even happen that one
well-chosen time-dependent basis state is enough to describe the dynamics of
the system very accurately. The time-dependent energy density functional
treatment of the giant resonances in nuclear physics provides such an
example~\cite{scampsSystematics2013,scampsSystematic2014}.

The second drawback of the TDGCM is the construction of a family of generator
states before the determination of the system evolution. The equation of motion
provides only the probability of the system to populate parts of this
predefined space. For this approach to work, the physicist must rely on an
\textit{a priori} knowledge of the relevant states for the dynamics. For
nuclear reactions, it typically means that one should correctly guess what will
be the reaction's output channels and include an ensemble of states
representative of these channels in the working space. Beyond the difficulty to
generate states representative of the systems far from the initial state, the
typical risks of this method are
\begin{itemize}
  \item to miss important channels/states in the construction of the set of
    generator states,
  \item to include states that will not be populated at all but will still
    increase the numerical cost.
\end{itemize}

\begin{figure}[ht!]
  \centering
  \includegraphics[width=0.60\linewidth]{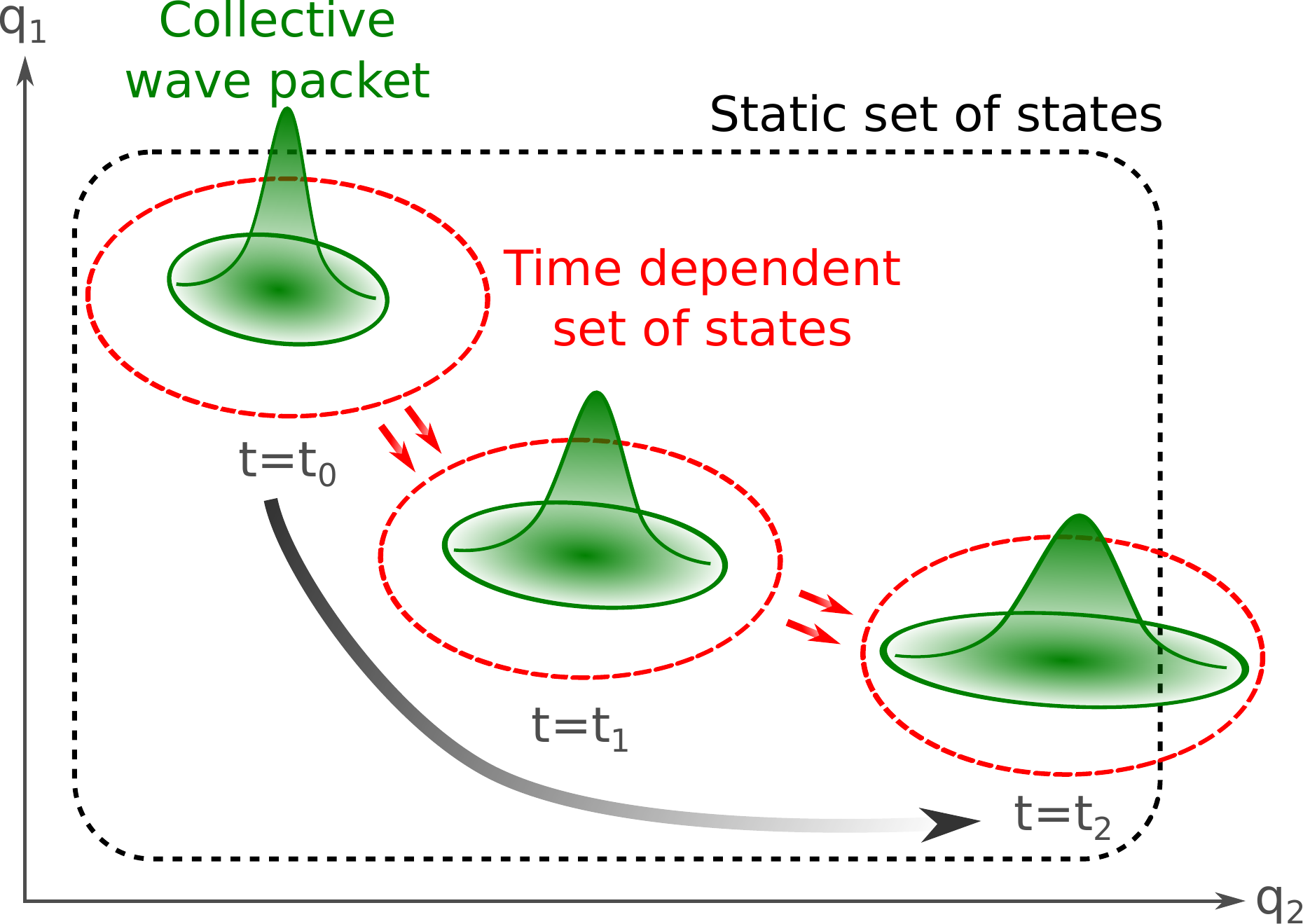}
  \caption{Schematic dynamics of a localized collective wave packet. With the
    TDGCM strategy (static set of generator states), most of the working space
    does not contribute, and part of the final state is not captured. In
    contrast, a small, well-chosen set of time-dependent states is sufficient
    to capture the system's dynamics entirely.}
    \label{fig:gcm_space}
\end{figure}

A solution to overcome these difficulties is the expansion of the
ansatz~\eqref{eq:gcm:GHW} on a set of time-dependent states, as shown
schematically in Fig.~\ref{fig:gcm_space}. In this case, the many-body wave function of the
system reads
\begin{equation}\label{eq:tdgcmtd}
  \ket{\Psi(t)} =
    \int_{\fvec{q} \in E} f(\fvec{q},t) \kHFB{\fvec{q},t}.
\end{equation}
This very general ansatz brings more flexibility as the configuration basis can
vary in time. However, this flexibility comes with additional complexity
in the equation of motion for the collective wavefunction $g(\fvec{q},t)$ and
the generator states $\kHFB{\fvec{q},t}$. Studies in
both chemistry and nuclear physics are exploring different strategies in the
choice of generator states and the determination of the equation of motion of
the system. We review these recent efforts in this Section.

\subsection{The multiconfiguration time-dependent Hartree-Fock approach}
\label{sec:mctdhf}

In 1990, Meyer \textit{et al.} introduced the multiconfiguration time-dependent
Hartree (MCTDH) approach to tackle the dynamics of
molecules~\cite{meyerMulticonfigurational1990}. Contrary to the fermionic
many-body problem, the system's degrees of freedom are distinct from each other
and correspond typically to distances between some atoms of a molecule. Their
associated wave function is assumed to be at any time a mixture of product
states
\begin{equation}
  \ket{\Psi(x_1,\cdots,x_n,t)} =
    \sum_{i_1=0}^{m_1-1}\cdots \sum_{i_n=0}^{m_n-1}
      c_{i_1\cdots i_n}
        \times \ket{\varphi_{i_1}^{(1)}(t)}\cdots\ket{\varphi_{i_n}^{(n)}(t)},
\end{equation}
where at any time, the $\{|\varphi_{i_k}^{(k)}(t)\}$ form a basis of the space
associated with the $k$th degree of freedom, and the $c_{i_1\cdots i_n}$ are
the mixing coefficients between all the product states. The equation of motion
of both the individual states and the mixing coefficients can then be obtained
from applying the Dirac-Frenkel variational principle. This method was since
applied to different dynamical processes in chemistry~\cite{mantheWave1992,
beckMulticonfiguration2000,meyerMultidimensional2009} and up to five degrees of
freedom in the treatment of the inelastic cross-section of H$_2$O +
H$_2$~\cite{ndengueStatetostate2019}. Note that in 2003, Wang \textit{et al.}
proposed an extension of this method referred to as multilayers MCTDHF to
tackle more degrees of freedom (up to a few thousand)~\cite{wangMultilayer2003}.

A natural extension of this work to the fermionic many-body problem is the
replacement of the product states by Slater determinants in the trial
wavefunction. This extension was introduced in
Ref.~\cite{nestMulticonfiguration2005} and the new ansatz reads
\begin{equation}\label{eq:mctdh_trial}
  \ket{\Psi(\fvec{r}_1,\cdots,\fvec{r}_n,t)} =
    \sum_{i_1=0}^{m_1-1}\cdots \sum_{i_n=0}^{m_n-1}
      c_{i_1\cdots i_n} \ket{\phi_{i_1\cdots i_n}(t)},
\end{equation}
with the time-dependent Slater determinants
\begin{equation}
  \ket{\phi_{i_1\cdots i_n}(t)} =
        \cPw{i_1}(t) \cdots \cPw{i_n}(t) \ket{0}.
\end{equation}
In this expression, $\cPw{i_k}(t)$ stands for the fermionic creation operator
of a particle in a single-particle state $\ket{\varphi_{i_k}(\fvec{r},t)}$.
This many-body wave function can then be injected into a time-dependent
variational principle whose parameters are both the mixing coefficients
$c_{i_1\cdots i_n}$ and the single-particle wave functions
$\ket{\varphi_{i_k}}$. Note that there is no one-to-one mapping between the
many-body state $\ket{\psi(t)}$ and the parameters of the right-hand side. In
practical applications, the set of single-particle wave functions is assumed to
be orthonormal at any time
\begin{equation}
  \braket{\varphi_{i}(t)}{\varphi_{j}(t)} = \delta_{ij}.
\end{equation}
This criterion lets some freedom in the choice of the $c_{i_k}$ and
$\ket{\varphi_{i_k}}$ for a given many-body wave function, leading to an
additional degree of freedom in their associated equation of motion. A usual
convention to fix this freedom consists in imposing the additional constraint
\begin{equation}
  \braket[\frac{\partial}{\partial t}]{\varphi_{i}(t)}{\varphi_{j}(t)} = 0.
\end{equation}
This choice stabilizes the single-particle states against rotations among the
occupied states. If such rotation has to be described, only the mixing
coefficients will be affected while the single-particle states will stay
constant. This convention yields to equations of motion that are often more
suited for the numerical time integration.

With this criterion, the Dirac-Frenkel variational principle applied to a
two-body Hamiltonian system leads to the equation of motion for both the
coefficients and the single-particle states
\begin{align}
  i \hbar \, \dot{c}_{i_1\cdots i_n} (t) &=
     \sum_{i_1=0}^{m_1-1}\cdots \sum_{i_n=0}^{m_n-1}
    \braket[\hat{H}]{\phi_{i_1\cdots i_n}(t)}{\phi_{i_1\cdots i_n}(t)} c_{i_1\cdots i_m}(t) \\
i\hbar \ket{\dot{\varphi}_n(t)} &=
    \hat{P} \left\{
      \hat{t} \ket{\varphi_n(t)}
    + \sum_{pqrs} (\rho^{-1})_{np}\,  \rho^{(2)}_{qspr} \, 
    \hat{h}_{rs} \ket{\varphi_q(t)}
    \right\}
\end{align}
where $\hat{t}$ is the one-body part of the Hamiltonian, $\hat{h}$ is the
mean-field potential that implicitly depends on the one-body density, $\rho$
and $\rho^{(2)}$ are the one- and two-body density matrices and $\hat{P}$ is a
projection operator on the orthogonal complement of the occupied
single-particle states. Such equation of motions have then been numerically
solved for chemical systems with six valence
electrons~\cite{nestMulticonfiguration2005}, to study the two photons
ionization of helium~\cite{hochstuhlTwophoton2011} or the dynamics of
di-molecular molecules~\cite{katoTime2004,katoTime2008}. In nuclear physics,
the multiconfiguration Hartree-Fock approach has been applied in its static
version to determine the structure of light nuclei mostly in the s-d
shell~\cite{robinDescription2016,robinDescription2017}. Such an expansion of
the many-body state enables a good description of the low lying excitation
spectrum with typically the first $2^+$ excitation reproduced within a few 100
keV. For the ground-state binding energy, this work still emphasizes a
significant overestimation of the theory by 8.3 MeV in average in the s-d shell
nuclei. This discrepancy would mostly come from (i) double-counting coming from
the usage of energy functional that have been fitted at the mean-field level,
(ii) the truncation of the configuration space that still cuts too early the
population of single-particle states with the largest spatial expansion.
Even though it would be interesting to study photoabsorption phenomena in
light nuclei or diffusion between light nuclei, this method has not yet been
applied in its dynamics version for nuclear physics. A generalization of the
ansatz~\eqref{eq:mctdh_trial} to a superposition of Bogoliubov vacua and its
corresponding equation of motion is yet to be formalized and tested.

\subsection{Multiconfiguration with time-dependent non orthogonal states}
\label{mctdhfi}

The trial state of Eq.~\eqref{eq:mctdh_trial} at the core of the MCTDHF method
expands the wave function on a set of orthonormal Slater determinants. The
orthonormality between such generator states simplifies the equation of motion
as typically the norm kernel defined in Eq.~\eqref{eq:gcm:nhker} is the
identity at any time. In contrast, it may be more efficient in some situations
to expand the many-body wave function on a set of non-orthogonal generator
states (i.e., time-dependent Bogoliubov vacua with time-dependent
deformations). Such a strategy was explored, for instance, in chemistry by
mixing TDDFT trajectories with a shift in time to include memory
effect~\cite{orestesGenerator2007} into the dynamics. This approach was proven
to correctly include the description of dissipation in the two electrons
dynamics of a Hooke's atom.

\begin{figure}[ht!]
  \begin{tabular}{cc}
    \begin{minipage}{0.52\textwidth}
      \includegraphics[width=1.\linewidth]{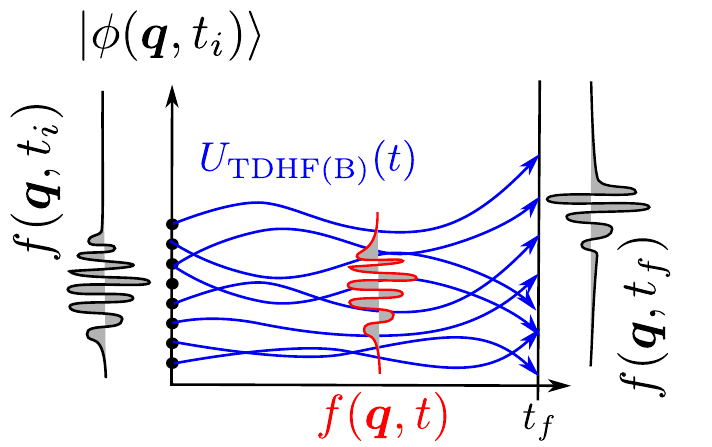} 
    \end{minipage} 
    &
    \begin{minipage}{0.42\textwidth}
      \includegraphics[width=1.\linewidth]{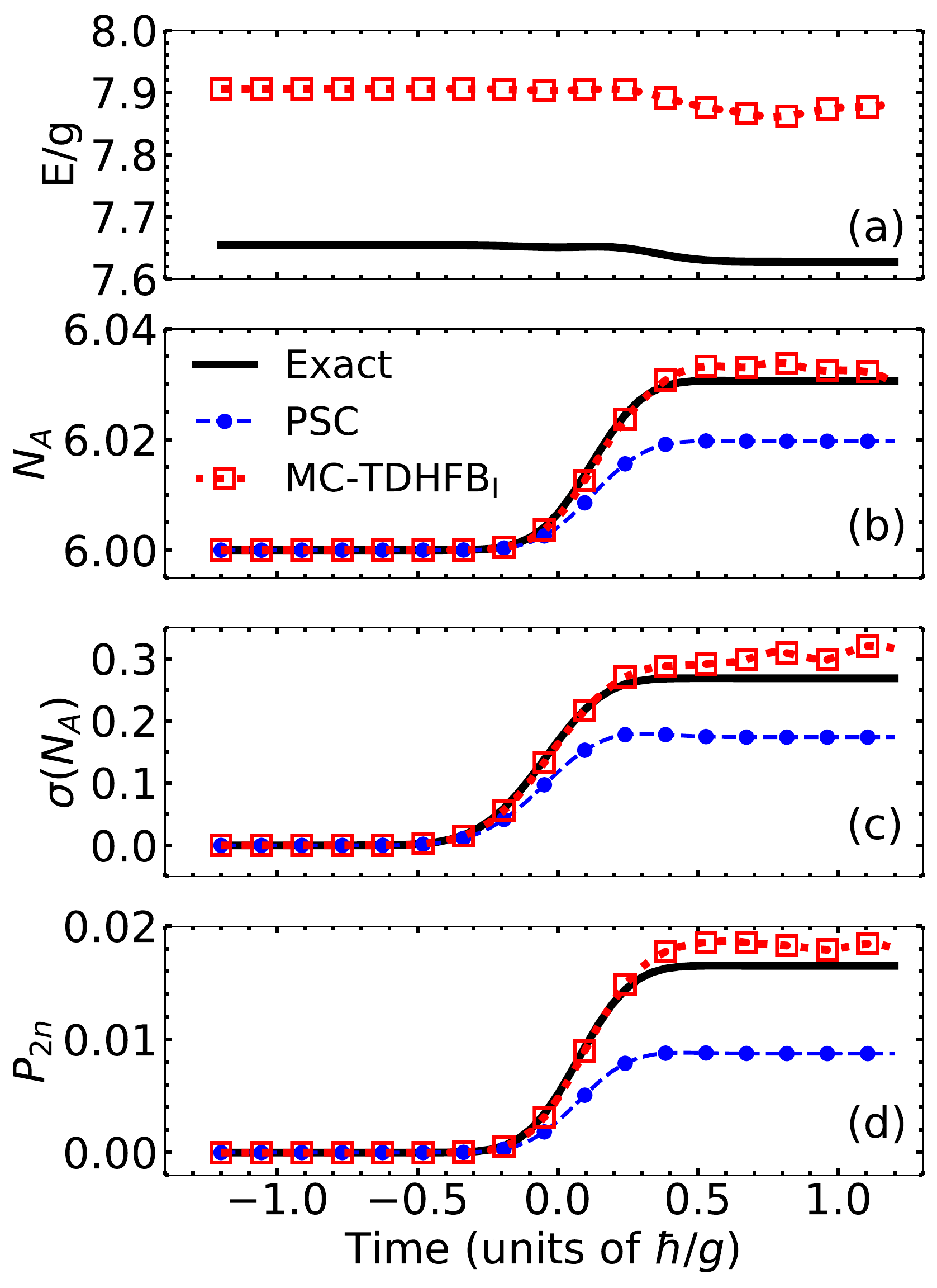}
    \end{minipage} 
  \end{tabular}
  \caption{Left panel: Schematic picture of a GCM mixing of TDHF(B) trajectories.
  Right panel: 
  Evolution of several observables during the dynamics of a simple model of collision between
  two superfluids.
  The total energy (a), the number of particles in the left subsystem (b) and its dispersion (c)
  as well as the probability to transfer one pair of fermions during the collision (d) are 
  plotted against the time of the collision for the exact many-body solution (black line), a
  statistical mixture of TDHFB trajectories (blue circles) and a quantum mixture of the same
  TDHFB trajectories (red squares).
  Reprinted figure with permission from Ref.~\cite{regnierMicroscopic2019}.
  Copyright 2019 by the American Physical Society.
  }
  \label{fig:mctdhfi}
\end{figure}

In nuclear physics, the idea of mixing time-dependent TDHF
trajectories was already proposed in 1983 in the pioneering work of
Reinhard \textit{et al.}~\cite{reinhardTime1983} to treat nuclear
collisions. Starting back from the ansatz~\eqref{eq:mctdh_trial}, the
authors proposed to take as the time-dependent generator states a set of
TDHF trajectories starting from different initial conditions. A time-dependent
variational principle is then applied to obtain the equation of motion only for
the mixing function $f(\fvec{q},t)$ (or the collective wave function
$g(\fvec{q},t)$). Such a principle is schematically pictured in
Fig.~\ref{fig:mctdhfi} (left panel).
The idea behind this scheme is that the TDHF trajectories will carry most of the
one-body dynamics of the system, whereas the weight function will encompass part
of the two-body collisional dynamics, in such importance as to account for additional
dissipation and fluctuation. In this paper, a GOA approximation was performed to
determine the evolution of the collective wave function in a one-dimensional
nuclear collision model. The results showed in particular that the widths of the
internal excitation energy of the collision partners after the collision were
increased by a factor of 7 compared to a TDHF trajectory alone. This additional
fluctuation is directly coming from the additional correlations tackled by the
enriched ansatz for the many-body wave function.

Even though promising, applications of this method to realistic systems were not
carried out. One possible explanation is the numerical cost associated
with TDHF trajectories. However, the advances in numerical methods and the
recent development of supercomputers induced a surge of interest for such
studies. In particular, the inclusion of superfluidity in our time-dependent
mean-field codes~\cite{scampsPairing2012,bulgacInduced2016} opened the
possibility to predict collisions between open-shell nuclei. Along this line,
Scamps~\textit{et al.} attempted to predict the transfer of pairs of fermions
in the contact between two superfluids based on a statistical mixing of TDHFB
trajectories~\cite{scampsSuperfluid2017,scampsTransfer2017,
regnierMicroscopic2018}. The idea is that the one-body dynamics of the nuclear
processes would be already well accounted for by TDHFB like trajectories, while
a statistical ensemble of such trajectories would account for the additional
fluctuation induced by the residual two-body collisions terms of the dynamics.
Up to now, these methods were only tested on toy-model cases and collisions
between a few light systems such as $^{20}$O+$^{20}$O. Experimental data on
such collisions still lack, which prevents a rigorous theory versus experiment
comparison.

Nevertheless, the tests on exactly solvable models show that these
semiclassical approaches manage to recover some crucial fluctuation related to
the relative gauge angle between the reaction partners. In particular, they can
predict the probability of one pair transfer with the proper order of magnitude
in the perturbative regime where the nuclear interaction during the collision
is weak compared to the pairing forces acting in each subsystem. Still, they
partially miss the quantum interference between the TDHFB trajectories.
Depending on the method's details, this may either lead to underestimating
fluctuations of one-body observables or, in the worse case, predicting
unphysical behavior such as particle transfer after the re-separation of the
two reaction partners.

Coming back to a full quantum treatment of the problem, Regnier~\textit{et al.}
recently attempted the full-fledged mixing of TDHFB trajectories in
Ref.~\cite{regnierMicroscopic2019}. In this context, the time-dependent
variational principle on the ansatz~\ref{eq:mctdh_trial} leads to the
equation of motion of the collective wave function $g(\fvec{q},t)$
\begin{equation}
\begin{array}{rl}
 i\hbar \dot{g} &= \left(\col{H} -\col{D} + i\hbar \dot{\nker}^{1/2}\nker^{-1/2} \right)g.
\end{array}
\label{eq:tdhw_proj_g}
\end{equation}
This equation involves the collective operators $\col{H}$ and $\col{D}$ defined
by the application of Eq.~\ref{eq:colop} on the kernels
\begin{align}
  \hker(\fvec{q},\fvec{q}') &=
    \bHFB{\fvec{q},t} \hat{H} \kHFB{\fvec{q}',t}, \\
   \kerRep{D}(\fvec{q}, \fvec{q}') &=
    \bHFB{\fvec{q},t} i\hbar\frac{\partial}{\partial t} \kHFB{\fvec{q}',t}.
\end{align}
All the kernels and collective operators involved now depend on time. Compared
to the TDGCM on static generator states (Eq.~\eqref{eq:tdgcm_coll}), this
equation contains two additional terms. The first one contains the time
derivative of the generator states, whereas the second one is linked to the
time derivative of the norm kernel. These equations were numerically solved
only in a simple case modeling the contact between two superfluid systems. The
main results are summarized in the right panel of Fig.~\ref{fig:mctdhfi}. The
full black line represents the system's exact many-body dynamics, and it is
compared with a prediction obtained from a statistical mixture of TDHFB
trajectories (the PSC method, dashed blue line) as well as the quantum mixing of the
same TDHFB trajectories. While the statistical method recovers the good order
of magnitude for most predictions, the inclusion of interference between the
TDHFB trajectories significantly improves these results. In particular, a
factor of two is highlighted between each method's predictions of the
probability $P_{2n}$ to transfer a pair of fermions during the collision.

At a time where performing series of independent time-dependent mean-field
calculations in nuclear physics becomes possible, such a method could be a
suitable candidate to tackle nuclear reactions with a complex interplay between
one-body and many-body degrees of freedom. The caveat to its direct application
on a realistic nuclear collision would still be the difficulty that current
implementations of the nuclear mean-field dynamics formalisms rely on energy
density functionals instead of a linear Hamiltonian (cf.
Sec.~\ref{sec:multireference}).

\section{Conclusion}
\label{sec:conclu}

This review presents four variants of the Time-Dependent Generator Coordinates
Method that is rooted in a configuration-mixing principle. This class of
methods is of particular interest to microscopically describe heavy-fermion
systems. It allows the physicist to focus the description on the correlations
of interest through the choice of the collective coordinates. Most of the time,
the collective coordinates are related to some of the first multipole moments
of the intrinsic one-body density or some groups of symmetry operators. Such
freedom makes the TDGCM extremely versatile. Still, its practical
applications in nuclear physics are plagued by the usage of effective
Hamiltonians or energy density functionals that lead to misbehaviors of the
energy kernel, an essential ingredient shared by all the TDGCM approaches.

The Time-Dependent Generator Coordinate Method is the most direct implementation
of the configuration-mixing principle. In this case, the only approximations are
the expression of the nuclear Hamiltonian, and the restriction of the total Fock
space to the one spanned by the configuration basis. The Griffin-Hill-Wheeler
(GHW) equation~\eqref{eq:gcm:GHW} is the corresponding equation of motion. The
main limitation of this method arises from the non-locality of the GHW equation
in the collective coordinates representation, leading to intensive parallel
computation. By resorting to some approximations, it is possible to rewrite the
GHW equations into local equations, reducing hereafter substantially the
calculation needs. The Gaussian Overlap Approximation (GOA) transforms the GHW
equations to a local Schrodinger-like equation essentially under the condition
that the norm kernel is of Gaussian character. The TDGCM+GOA is the most widely
used implementation of the TDGCM for the description of nuclear reactions and
especially for fission. The Schrodinger Collective-Intrinsic Method is based on
the truncation of the GHW equation in the second-order to obtain a local
Schrodinger-like equation. In its current form, it still requires to calculate
the full norm and Hamiltonian kernels. Finally, it is possible to generalize
the standard TDGCM approach by expanding the many-body wave function on a set
of time-dependent generator states. The recent progress of TDHFB solvers opens
new possibilities for practical applications along this line.

Overall, most of these methods were first developed in the 1980s, at a time when
they were quickly facing intractable numerical costs. The computational power at
our disposal nowadays is an incentive to revisit the TDGCM approaches and look
for new opportunities in the description of nuclear reactions. One of the most
significant challenges in this path is the determination of energy density
functionals or effective Hamiltonians, which are compatible with the GCM
formalism and yield quantitative predictions of nuclear observables.

\section{Aknowledgements}
\label{sec:aknowledgements}
The authors would like to thank N. Schunck for his patience through the writing
process of this paper as well as for fruitful comments on the manuscript.
This work was performed at Los Alamos National Laboratory under the auspices of
the National Nuclear Security Administration of the U.S. Department of Energy at
Los Alamos National Laboratory under Contract No. 89233218CNA000001. Support for
this work was provided through the Fission In R-process Elements (FIRE) Topical
Collaboration in Nuclear Theory of the U.S. Department of Energy. It was partly
performed under the auspices of the U.S. Department of Energy by the Lawrence
Livermore National Laboratory (LLNL) under Contract No. DE-AC52-07NA27344.

\appendix
\section{Expression of the GOA moments}
\label{app:goa_moments}

The expression~\eqref{eq:goa_gb_moments} involves some moments $M^{(K)}$
and $\tilde{M}^{(K)}$ that we define here. We recall that we consider generator
states that are Bogoliubov vacua. Any generator state is then fully
characterized by its generalized density matrix $\mathcal{R}(\fvec{q})$
\begin{equation}
 \mathcal{R} =
 \left[
 \begin{array}{cc}
  \rho  & \kappa \\
  -\kappa^ * & 1-\rho^* 
 \end{array}
 \right].
\end{equation}
Additionally, each collective coordinate $\fvec{q}_i$ is associated with a
one-body observable $\op{Q}_i$ that is used as a constraint. Expressed in the
basis of quasiparticles that diagonalizes $\mathcal{R}(\fvec{q})$, this
operator takes the matrix form
\begin{equation}
 \mathcal{Q}_i =
 \left[
 \begin{array}{cc}
  Q^{11}_i  & Q^{12}_i \\
  Q^{21}_i  & Q^{22}_i 
 \end{array}
 \right].
\end{equation}
One can define the standard QRPA matrix $\mathcal{M}$ in this same basis as
detailed in Ref.~\cite{ringNuclear2004}. With these notations, the moments
$M^{(K)}$ involved in the determination of the GOA inertia and metric tensors
are
\begin{equation}
M_{ij}^{(K)} = \frac{1}{2} ( Q_{i}^{12\,*},\ Q_{i}^{12}) \mathcal{M}^{-K} \left(\begin{array}{c} Q_{j}^{12} \\ Q_{j}^{12\,*} \end{array}\right).
\label{eq:moments}
\end{equation}
We also define the modified moments $\tilde{M}^{(K)}$ by
\begin{equation}
 \tilde{M}^{(K)} = 
 \left[
 \begin{array}{cc}
  1   & 0 \\
  0 & -1
 \end{array}
 \right]
 M^{(K)}
 \left[
 \begin{array}{cc}
  1   & 0 \\
  0 & -1
 \end{array}
 \right].
\end{equation}

\section{Bestiary of SOPO properties}
\label{app:SOPO}
The Symmetric Ordered Product of Operators (SOPO) are defined, for any two
operators $\opRep{A}$ and $\opRep{B}$, as
\begin{equation}\label{eq:SCIM:SOPOdef}
  \SOPO{n}{\opRep{A}}{\opRep{B}} =
    \frac{1}{2^n}
      \sum_{k=0}^{n}
        \binom{n}{k}
        \opRep{B}^{k} \opRep{A} \opRep{B}^{n-k}.
\end{equation}
They can be equivalently defined recursively through their relation with the
anti-commutator
\begin{align}
  \SOPO{1}{\opRep{A}}{\opRep{B}} &=
    \frac{1}{2}\left\{\opRep{A},\opRep{B}\right\} \\
  \SOPO{n+1}{\opRep{A}}{\opRep{B}} &=
    \frac{1}{2}\left\{\SOPO{n}{\opRep{A}}{\opRep{B}},\opRep{B}\right\}.
\end{align}
The SOPO are used to obtain the Symmetric Moment Expansion (SME) of the
symmetrized GHW equation~\eqref{eq:SCIM:HWsym2}, based on
\begin{equation}\label{eq:SCIM:SOPOExpdecompo}
  e^{\alpha\opRep{B}/2}\opRep{A}e^{\alpha\opRep{B}/2}
   = \sum_{p=0}^{\infty} \frac{\alpha^p}{p!} \SOPO{p}{\opRep{A}}{\opRep{B}},
\end{equation}
For any operators $\opRep{A}$, $\opRep{B}$ and $\opRep{C}$, the following
relation is satisfied
\begin{equation}
  \opRep{A}\SOPO{n}{\opRep{B}}{\opRep{P}}\opRep{C} =
    \sum_{k = 0}^{n}\SOPO{k}{\opRep{B}^{A,C}_{(n,k)}}{\opRep{P}},
\end{equation}
where the operators $\opRep{B}^{A,C}_{(n,k)}$ are given by
\begin{equation}\label{eq:SCIM:SOPOsandwich}
  \opRep{B}^{A,C}_{(n,k)} =
    \frac{i^{n-k}}{2^{n-k}}
    \sum_{r=0}^{n-k}
      \Bigg\{
        (-1)^{r}
        \binom{n}{k+r}
        \binom{k+r}{r}
        \opRep{A}^{[r]}\ \opRep{B}\ \opRep{C}^{[n-k-r]}
      \Bigg\},
\end{equation}
and where $\opRep{A}^{[r]}$ the short-hand notation for the local operator
associated with the kernel
\begin{equation}
  \kerRep{A}^{[r]}(q) = \frac{\partial^r \kerRep{A}}{\partial q^r}(q).
\end{equation}

\bibliographystyle{IEEEtran}
\bibliography{biblio}

\begin{thebibliography}{100}
\providecommand{\url}[1]{#1}
\csname url@samestyle\endcsname
\providecommand{\newblock}{\relax}
\providecommand{\bibinfo}[2]{#2}
\providecommand{\BIBentrySTDinterwordspacing}{\spaceskip=0pt\relax}
\providecommand{\BIBentryALTinterwordstretchfactor}{4}
\providecommand{\BIBentryALTinterwordspacing}{\spaceskip=\fontdimen2\font plus
\BIBentryALTinterwordstretchfactor\fontdimen3\font minus
  \fontdimen4\font\relax}
\providecommand{\BIBforeignlanguage}[2]{{%
\expandafter\ifx\csname l@#1\endcsname\relax
\typeout{** WARNING: IEEEtran.bst: No hyphenation pattern has been}%
\typeout{** loaded for the language `#1'. Using the pattern for}%
\typeout{** the default language instead.}%
\else
\language=\csname l@#1\endcsname
\fi
#2}}
\providecommand{\BIBdecl}{\relax}
\BIBdecl

\bibitem{bohrMechanism1939}
N.~Bohr and J.~A. Wheeler, ``The {{Mechanism}} of {{Nuclear Fission}},''
  \emph{Phys. Rev.}, vol.~56, no.~5, pp. 426--450, Sep. 1939.

\bibitem{bohrNuclear1998}
A.~Bohr and B.~R. Mottelson, \emph{Nuclear Structure. 1, 1,}.\hskip 1em plus
  0.5em minus 0.4em\relax {Singapore; River Edge, NJ}: {World Scientific},
  1998.

\bibitem{scheidTheory1968}
W.~Scheid and W.~Greiner, ``Theory of projection of spurious center of mass and
  rotational states from many-body nuclear wave functions,'' \emph{Annals of
  Physics}, vol.~48, no.~3, pp. 493--525, Jul. 1968.

\bibitem{finkSpurious1972}
B.~Fink, D.~Kolb, W.~Scheid, and W.~Greiner, ``Spurious rotational states in
  deformed nuclear shell models,'' \emph{Annals of Physics}, vol.~69, no.~2,
  pp. 375--399, Feb. 1972.

\bibitem{hillNuclear1953}
D.~L. Hill and J.~A. Wheeler, ``Nuclear {{Constitution}} and the
  {{Interpretation}} of {{Fission Phenomena}},'' \emph{Phys. Rev.}, vol.~89,
  no.~5, pp. 1102--1145, Mar. 1953.

\bibitem{griffinCollective1957}
J.~J. Griffin and J.~A. Wheeler, ``Collective {{Motions}} in {{Nuclei}} by the
  {{Method}} of {{Generator Coordinates}},'' \emph{Phys. Rev.}, vol. 108,
  no.~2, pp. 311--327, Oct. 1957.

\bibitem{benderGoing2008}
M.~Bender, ``Going beyond the self-consistent mean-field with the
  symmetry-restored generator coordinate method: {{From}} single-reference
  toward multi-reference energy density functional theory,'' \emph{Eur. Phys.
  J. Spec. Top.}, vol. 156, no.~1, pp. 217--228, Apr. 2008.

\bibitem{egidoStateoftheart2016}
J.~L. Egido, ``State-of-the-art of beyond mean field theories with nuclear
  density functionals,'' \emph{Phys. Scr.}, vol.~91, no.~7, p. 073003, Jun.
  2016.

\bibitem{krewaldSelfconsistent1976}
\BIBentryALTinterwordspacing
S.~Krewald, R.~Rosenfelder, J.~Galonska, and A.~Faessler, ``Selfconsistent
  generator coordinate method for giant monopole resonances,'' \emph{Nuclear
  Physics A}, vol. 269, no.~1, pp. 112 -- 124, 1976. [Online]. Available:
  \url{http://www.sciencedirect.com/science/article/pii/0375947476904000}
\BIBentrySTDinterwordspacing

\bibitem{stoitsovGenerator1994}
\BIBentryALTinterwordspacing
M.~V. Stoitsov, P.~Ring, and M.~M. Sharma, ``Generator coordinate calculations
  for breathing-mode giant monopole resonance in the relativistic mean-field
  theory,'' \emph{Phys. Rev. C}, vol.~50, pp. 1445--1455, Sep 1994. [Online].
  Available: \url{https://link.aps.org/doi/10.1103/PhysRevC.50.1445}
\BIBentrySTDinterwordspacing

\bibitem{caurierMicroscopic1973}
\BIBentryALTinterwordspacing
E.~Caurier, B.~Bourotte-Bilwes, and Y.~Abgrall, ``Microscopic treatment of the
  coupled monopole and quadrupole vibrations in light nuclei,'' \emph{Physics
  Letters B}, vol.~44, no.~5, pp. 411 -- 415, 1973. [Online]. Available:
  \url{http://www.sciencedirect.com/science/article/pii/0370269373903213}
\BIBentrySTDinterwordspacing

\bibitem{flocardGenerator1976}
\BIBentryALTinterwordspacing
H.~Flocard and D.~Vautherin, ``Generator coordinate calculations of giant
  resonances with the skyrme interaction,'' \emph{Nuclear Physics A}, vol. 264,
  no.~2, pp. 197 -- 220, 1976. [Online]. Available:
  \url{http://www.sciencedirect.com/science/article/pii/0375947476904280}
\BIBentrySTDinterwordspacing

\bibitem{abgrallMonopole1975}
\BIBentryALTinterwordspacing
Y.~Abgrall and E.~Caurier, ``On the monopole and quadrupole isoscalar giant
  resonances in 4he, 16o, 20ne and 40ca,'' \emph{Physics Letters B}, vol.~56,
  no.~3, pp. 229 -- 231, 1975. [Online]. Available:
  \url{http://www.sciencedirect.com/science/article/pii/0370269375903810}
\BIBentrySTDinterwordspacing

\bibitem{vretenarIsoscalar1997}
\BIBentryALTinterwordspacing
D.~Vretenar, G.~Lalazissis, R.~Behnsch, W.~Pöschl, and P.~Ring, ``Monopole
  giant resonances and nuclear compressibility in relativistic mean field
  theory,'' \emph{Nuclear Physics A}, vol. 621, no.~4, pp. 853 -- 878, 1997.
  [Online]. Available:
  \url{http://www.sciencedirect.com/science/article/pii/S0375947497001929}
\BIBentrySTDinterwordspacing

\bibitem{deKinematics1978}
A.~De~Toledo~Piza and E.~De~Passos, ``The kinematics of generator
  co-ordinates,'' \emph{Il Nuovo Cimento B (1971-1996)}, vol.~45, no.~1, pp.
  1--30, 1978.

\bibitem{ringNuclear2004}
P.~Ring and P.~Schuck, \emph{The {{Nuclear Many}}-{{Body Problem}}}.\hskip 1em
  plus 0.5em minus 0.4em\relax {Springer Science \& Business Media}, Mar. 2004.

\bibitem{lowdinComments1972}
P.~O. L{\"o}wdin and P.~K. Mukherjee, ``Some comments on the time-dependent
  variation principle,'' \emph{Chemical Physics Letters}, vol.~14, no.~1, pp.
  1--7, May 1972.

\bibitem{crankPractical1996}
J.~Crank and P.~Nicolson, ``A practical method for numerical evaluation of
  solutions of partial differential equations of the heat-conduction type,''
  \emph{Adv. Comput. Math.}, vol.~6, no. 3-4, pp. 207--226, 1996.

\bibitem{reedMethods1972}
M.~Reed and B.~Simon, \emph{Methods of Modern Mathematical Physics}.\hskip 1em
  plus 0.5em minus 0.4em\relax Academic Press, 1972.

\bibitem{robledoSign2009}
L.~M. Robledo, ``Sign of the overlap of {{Hartree}}-{{Fock}}-{{Bogoliubov}}
  wave functions,'' \emph{Phys. Rev. C}, vol.~79, no.~2, p. 021302, Feb. 2009.

\bibitem{anguianoParticle2001}
M.~Anguiano, J.~L. Egido, and L.~M. Robledo, ``Particle number projection with
  effective forces,'' \emph{Nuclear Physics A}, vol. 696, no.~3, pp. 467--493,
  Dec. 2001.

\bibitem{lacroixConfiguration2009}
D.~Lacroix, T.~Duguet, and M.~Bender, ``Configuration mixing within the energy
  density functional formalism: {{Removing}} spurious contributions from
  nondiagonal energy kernels,'' \emph{Phys. Rev. C}, vol.~79, no.~4, Apr. 2009.

\bibitem{benderParticlenumber2009}
M.~Bender, T.~Duguet, and D.~Lacroix, ``Particle-number restoration within the
  energy density functional formalism,'' \emph{Phys. Rev. C}, vol.~79, no.~4,
  p. 044319, Apr. 2009.

\bibitem{duguetParticlenumber2009}
T.~Duguet, M.~Bender, K.~Bennaceur, D.~Lacroix, and T.~Lesinski,
  ``Particle-number restoration within the energy density functional formalism:
  {{Nonviability}} of terms depending on noninteger powers of the density
  matrices,'' \emph{Phys. Rev. C}, vol.~79, no.~4, p. 044320, Apr. 2009.

\bibitem{sheikhSymmetry2019}
J.~A. Sheikh, J.~Dobaczewski, P.~Ring, L.~M. Robledo, and C.~Yannouleas,
  ``Symmetry restoration in mean-field approaches,'' \emph{arXiv:1901.06992},
  2019.

\bibitem{ryssensSolution2015}
W.~Ryssens, V.~Hellemans, M.~Bender, and P.~H. Heenen, ``Solution of the
  {{Skyrme}}\textendash{{HF}}+{{BCS}} equation on a {{3D}} mesh, {{II}}: {{A}}
  new version of the {{Ev8}} code,'' \emph{Computer Physics Communications},
  vol. 187, pp. 175--194, Feb. 2015.

\bibitem{donauCanonical1998}
F.~D{\"o}nau, ``Canonical form of transition matrix elements,'' \emph{Phys.
  Rev. C}, vol.~58, no.~2, pp. 872--877, Aug. 1998.

\bibitem{anguianoCoulomb2001}
M.~Anguiano, J.~L. Egido, and L.~M. Robledo, ``Coulomb exchange and pairing
  contributions in nuclear
  {{Hartree}}\textendash{{Fock}}\textendash{{Bogoliubov}} calculations with the
  {{Gogny}} force,'' \emph{Nuclear Physics A}, vol. 683, no.~1, pp. 227--254,
  Feb. 2001.

\bibitem{almehedPairing2001}
D.~Almehed, S.~Frauendorf, and F.~D{\"o}nau, ``Pairing correlations in
  high-{{K}} bands,'' \emph{Phys. Rev. C}, vol.~63, no.~4, p. 044311, Mar.
  2001.

\bibitem{dobaczewskiParticlenumber2007}
J.~Dobaczewski, M.~V. Stoitsov, W.~Nazarewicz, and P.-G. Reinhard,
  ``Particle-number projection and the density functional theory,'' \emph{Phys.
  Rev. C}, vol.~76, no.~5, p. 054315, Nov. 2007.

\bibitem{duguetDensity2003}
T.~Duguet and P.~Bonche, ``Density dependence of two-body interactions for
  beyond--mean-field calculations,'' \emph{Phys. Rev. C}, vol.~67, no.~5, p.
  054308, May 2003.

\bibitem{robledoRemarks2010}
L.~M. Robledo, ``Remarks on the use of projected densities in the
  density-dependent part of {{Skyrme}} or {{Gogny}} functionals,'' \emph{J.
  Phys. G: Nucl. Part. Phys.}, vol.~37, no.~6, p. 064020, Apr. 2010.

\bibitem{rodriguez-guzmanCorrelations2002}
R.~{Rodr{\'i}guez-Guzm{\'a}n}, J.~L. Egido, and L.~M. Robledo, ``Correlations
  beyond the mean field in magnesium isotopes: Angular momentum projection and
  configuration mixing,'' \emph{Nuclear Physics A}, vol. 709, no.~1, pp.
  201--235, Oct. 2002.

\bibitem{bennaceurNonlocal2017}
K.~Bennaceur, A.~Idini, J.~Dobaczewski, P.~Dobaczewski, M.~Kortelainen, and
  F.~Raimondi, ``Nonlocal energy density functionals for pairing and
  beyond-mean-field calculations,'' \emph{J. Phys. G: Nucl. Part. Phys.},
  vol.~44, no.~4, p. 045106, 2017.

\bibitem{yaoInitio2019}
J.~M. Yao, B.~Bally, J.~Engel, R.~Wirth, T.~R. Rodr{\'i}guez, and H.~Hergert,
  ``Ab {{Initio Treatment}} of {{Collective Correlations}} and the
  {{Neutrinoless Double Beta Decay}} of 48ca,'' \emph{arXiv:1908.05424}, Aug.
  2019.

\bibitem{verriereDescription2017}
\BIBentryALTinterwordspacing
M.~Verriere, ``{Description de la dynamique de la fission dans le formalisme de
  la m{\'e}thode de la coordonn{\'e}e g{\'e}n{\'e}ratrice d{\'e}pendante du
  temps},'' Ph.D. dissertation, Universit{\'e} Paris-Saclay, May 2017.
  [Online]. Available:
  \url{https://tel.archives-ouvertes.fr/tel-01559158/document}
\BIBentrySTDinterwordspacing

\bibitem{verriereFission2017}
M.~Verriere, N.~Dubray, N.~Schunck, D.~Regnier, and P.~{Dossantos-Uzarralde},
  ``Fission description: {{First}} steps towards a full resolution of the
  time-dependent {{Hill}}-{{Wheeler}} equation,'' \emph{EPJ Web Conf.}, vol.
  146, p. 04034, 2017.

\bibitem{dubrayNumerical2012}
N.~Dubray and D.~Regnier, ``Numerical search of discontinuities in
  self-consistent potential energy surfaces,'' \emph{Comput. Phys. Commun.},
  vol. 183, no.~10, pp. 2035--2041, Oct. 2012.

\bibitem{nishioMeasurement1995}
\BIBentryALTinterwordspacing
K.~NISHIO, Y.~NAKAGOME, I.~KANNO, and I.~KIMURA, ``Measurement of fragment mass
  dependent kinetic energy and neutron multiplicity for thermal neutron induced
  fission of plutonium-239,'' \emph{Journal of Nuclear Science and Technology},
  vol.~32, no.~5, pp. 404--414, 1995. [Online]. Available:
  \url{https://doi.org/10.1080/18811248.1995.9731725}
\BIBentrySTDinterwordspacing

\bibitem{tsuchiyaSimultaneous2000}
\BIBentryALTinterwordspacing
C.~TSUCHIYA, Y.~NAKAGOME, H.~YAMANA, H.~MORIYAMA, K.~NISHIO, I.~KANNO, K.~SHIN,
  and I.~KIMURA, ``Simultaneous measurement of prompt neutrons and fission
  fragments for 239pu(nth,f),'' \emph{Journal of Nuclear Science and
  Technology}, vol.~37, no.~11, pp. 941--948, 2000. [Online]. Available:
  \url{https://doi.org/10.1080/18811248.2000.9714976}
\BIBentrySTDinterwordspacing

\bibitem{goutteMicroscopic2005}
H.~Goutte, J.~F. Berger, P.~Casoli, and D.~Gogny, ``Microscopic approach of
  fission dynamics applied to fragment kinetic energy and mass distributions in
  {{U238}},'' \emph{Phys. Rev. C}, vol.~71, no.~2, p. 024316, Feb. 2005.

\bibitem{bohrCoupling1952}
A.~Bohr, ``The coupling of nuclear surface oscillations to the motion of
  individual nucleons,'' \emph{Dan. Mat. Fys. Medd.}, vol.~26, no.~14, pp.
  1--40, 1952.

\bibitem{kumarComplete1967}
K.~Kumar and M.~Baranger, ``Complete numerical solution of {{Bohr}}'s
  collective {{Hamiltonian}},'' \emph{Nuclear Physics A}, vol.~92, no.~3, pp.
  608--652, Feb. 1967.

\bibitem{libertMicroscopic1999}
J.~Libert, M.~Girod, and J.-P. Delaroche, ``Microscopic descriptions of
  superdeformed bands with the {{Gogny}} force: {{Configuration}} mixing
  calculations in the $a\simeq$190 mass region,'' \emph{Phys. Rev. C}, vol.~60,
  no.~5, p. 054301, Sep. 1999.

\bibitem{delarocheStructure2010}
J.~P. Delaroche, M.~Girod, J.~Libert, H.~Goutte, S.~Hilaire, S.~P{\'e}ru,
  N.~Pillet, and G.~F. Bertsch, ``Structure of even-even nuclei using a mapped
  collective {{Hamiltonian}} and the {{D1S Gogny}} interaction,'' \emph{Phys.
  Rev. C}, vol.~81, no.~1, p. 014303, Jan. 2010.

\bibitem{fuBeyond2013}
\BIBentryALTinterwordspacing
Y.~Fu, H.~Mei, J.~Xiang, Z.~P. Li, J.~M. Yao, and J.~Meng, ``Beyond
  relativistic mean-field studies of low-lying states in neutron-deficient
  krypton isotopes,'' \emph{Phys. Rev. C}, vol.~87, p. 054305, May 2013.
  [Online]. Available:
  \url{https://link.aps.org/doi/10.1103/PhysRevC.87.054305}
\BIBentrySTDinterwordspacing

\bibitem{matsuyanagiMicroscopic2016}
K.~Matsuyanagi, M.~Matsuo, T.~Nakatsukasa, K.~Yoshida, N.~Hinohara, and
  K.~Sato, ``Microscopic derivation of the {{Bohr}}\textendash{{Mottelson}}
  collective {{Hamiltonian}} and its application to quadrupole shape
  dynamics,'' \emph{Phys. Scr.}, vol.~91, no.~6, p. 063014, May 2016.

\bibitem{reinhardGeneratorcoordinate1987}
P.~Reinhard and K.~Goeke, ``The {{Generator}}-{{Coordinate Method}} and
  {{Quantized Collective Motion}} in {{Nuclear Systems}},'' \emph{Rep. Prog.
  Phys.}, vol.~50, no.~1, pp. 1--64, Jan. 1987.

\bibitem{krappeTheory2012}
\BIBentryALTinterwordspacing
H.~J. Krappe and K.~Pomorski, \emph{Theory of {{Nuclear Fission}}}.\hskip 1em
  plus 0.5em minus 0.4em\relax {Springer}, 2012. [Online]. Available:
  \url{http://www.springer.com/us/book/9783642235146}
\BIBentrySTDinterwordspacing

\bibitem{onishiLocal1975}
N.~Onishi and T.~Une, ``Local {{Gaussian Approximation}} in the {{Generator
  Coordinate Method}},'' \emph{Prog. Theor. Phys.}, vol.~53, no.~2, pp.
  504--515, Jan. 1975.

\bibitem{gozdzExtended1985}
A.~Gozdz, ``An extended gaussian overlap approximation in the generator
  coordinate method,'' \emph{Phys. Lett. B}, vol. 152, no. 5-6, pp. 281--283,
  Mar. 1985.

\bibitem{schunckMicroscopic2016}
N.~Schunck and L.~M. Robledo, ``Microscopic theory of nuclear fission: A
  review,'' \emph{Rep. Prog. Phys.}, vol.~79, no.~11, p. 116301, Oct. 2016.

\bibitem{holzwarthFour1973}
G.~Holzwarth, ``Four approaches to the function of inertia in a solvable
  model,'' \emph{Nuclear Physics A}, vol. 207, no.~3, pp. 545--564, Jun. 1973.

\bibitem{gozdzMass1985}
A.~G{\'o}{\'z}d{\'z}, K.~Pomorski, M.~Brack, and E.~Werner, ``The mass
  parameters for the average mean-field potential,'' \emph{Nuclear Physics A},
  vol. 442, no.~1, pp. 26--49, Aug. 1985.

\bibitem{baranQuadrupole2011}
A.~Baran, J.~Sheikh, J.~Dobaczewski, W.~Nazarewicz, and A.~Staszczak,
  ``Quadrupole collective inertia in nuclear fission: Cranking approximation,''
  \emph{Physical Review C}, vol.~84, no.~5, p. 054321, 2011.

\bibitem{reinhardGeneration1978}
P.-G. Reinhard and K.~Goeke, ``The generation coordinate method with conjugate
  parameters and its relation to adiabatic time-dependent
  {{Hartree}}-{{Fock}},'' \emph{J. Phys. G: Nucl. Phys.}, vol.~4, no.~9, p.
  L245, Sep. 1978.

\bibitem{bergerSelf1980}
J.~Berger and D.~Gogny, ``A self-consistent microscopic approach to the 12c+
  12c reaction at low energy,'' \emph{Nuclear Physics A}, vol. 333, no.~2, pp.
  302--332, 1980.

\bibitem{bayeGenerator1979}
D.~Baye and Y.~Salmon, ``Generator-coordinate study of elastic 40ca+ 40ca
  scattering,'' \emph{Nuclear Physics A}, vol. 323, no. 2-3, pp. 521--539,
  1979.

\bibitem{friedrichElastic1976}
H.~Friedrich, K.~Langanke, and A.~Weiguny, ``Elastic scattering of {{16O}} on
  {{40Ca}} at backward angles,'' \emph{Physics Letters B}, vol.~63, no.~2, pp.
  125--128, Jul. 1976.

\bibitem{goekeThreedimensional1983}
\BIBentryALTinterwordspacing
K.~Goeke, F.~Grümmer, and P.-G. Reinhard, ``Three-dimensional nuclear dynamics
  in the quantized atdhf approach,'' \emph{Annals of Physics}, vol. 150, no.~2,
  pp. 504 -- 551, 1983. [Online]. Available:
  \url{http://www.sciencedirect.com/science/article/pii/0003491683900258}
\BIBentrySTDinterwordspacing

\bibitem{sekizawaTDHF2019}
K.~Sekizawa, ``{{TDHF Theory}} and {{Its Extensions}} for the {{Multinucleon
  Transfer Reaction}}: {{A Mini Review}},'' \emph{Front. Phys.}, vol.~7, 2019.

\bibitem{bergerMicroscopic1984}
J.~F. Berger, M.~Girod, and D.~Gogny, ``Microscopic analysis of collective
  dynamics in low energy fission,'' \emph{Nuclear Physics A}, vol. 428, pp.
  23--36, Oct. 1984.

\bibitem{bergerTimedependent1991}
------, ``Time-dependent quantum collective dynamics applied to nuclear
  fission,'' \emph{Computer Physics Communications}, vol.~63, no.
  1\textendash{}3, pp. 365--374, Feb. 1991.

\bibitem{younesFragment2012}
W.~Younes and D.~Gogny, ``Fragment {{Yields Calculated}} in a
  {{Time}}-{{Dependent Microscopic Theory}} of {{Fission}},'' Lawrence
  Livermore National Laboratory, Livermore, CA, Tech. Rep. LLNL-TR-586678,
  2012.

\bibitem{younesCollective2012}
------, ``Collective {{Dissipation}} from {{Saddle}} to {{Scission}} in a
  {{Microscopic Approach}},'' Lawrence Livermore National Laboratory,
  Livermore, CA, Tech. Rep. LLNL-TR-586694, 2012.

\bibitem{regnierFelix12016}
D.~Regnier, M.~Verri{\`e}re, N.~Dubray, and N.~Schunck, ``{{FELIX}}-1.0: {{A}}
  finite element solver for the time dependent generator coordinate method with
  the {{Gaussian}} overlap approximation,'' \emph{Computer Physics
  Communications}, vol. 200, pp. 350--363, Mar. 2016.

\bibitem{regnierFelix22018}
D.~Regnier, N.~Dubray, M.~Verri{\`e}re, and N.~Schunck, ``{{FELIX}}-2.0:
  {{New}} version of the finite element solver for the time dependent generator
  coordinate method with the {{Gaussian}} overlap approximation,''
  \emph{Computer Physics Communications}, vol. 225, pp. 180--191, Apr. 2018.

\bibitem{regnierFission2016}
D.~Regnier, N.~Dubray, N.~Schunck, and M.~Verri{\`e}re, ``Fission fragment
  charge and mass distributions in {\textsuperscript{239}}{{Pu}}(n,f) in the
  adiabatic nuclear energy density functional theory,'' \emph{Phys. Rev. C},
  vol.~93, no.~5, p. 054611, May 2016.

\bibitem{taoMicroscopic2017}
H.~Tao, J.~Zhao, Z.~P. Li, T.~Nik{\v s}i{\'c}, and D.~Vretenar, ``Microscopic
  study of induced fission dynamics of {{Th}} 226 with covariant energy density
  functionals,'' \emph{Phys. Rev. C}, vol.~96, no.~2, Aug. 2017.

\bibitem{regnierAsymmetric2019}
D.~Regnier, N.~Dubray, and N.~Schunck, ``From asymmetric to symmetric fission
  in the fermium isotopes within the time-dependent generator-coordinate-method
  formalism,'' \emph{Phys. Rev. C}, vol.~99, no.~2, p. 024611, Feb. 2019.

\bibitem{zdebFission2017}
A.~Zdeb, A.~Dobrowolski, and M.~Warda, ``Fission dynamics of
  {\textsuperscript{252}}{{Cf}},'' \emph{Phys. Rev. C}, vol.~95, no.~5, p.
  054608, 2017.

\bibitem{zhaoMicroscopic2019}
J.~Zhao, T.~Nik{\v s}i{\'c}, D.~Vretenar, and S.-G. Zhou, ``Microscopic
  self-consistent description of induced fission dynamics:
  {{Finite}}-temperature effects,'' \emph{Phys. Rev. C}, vol.~99, no.~1, p.
  014618, Jan. 2019.

\bibitem{zhaoTimedependent2019}
J.~Zhao, J.~Xiang, Z.-P. Li, T.~Nik{\v s}i{\'c}, D.~Vretenar, and S.-G. Zhou,
  ``Time-dependent generator-coordinate-method study of mass-asymmetric fission
  of actinides,'' \emph{Phys. Rev. C}, vol.~99, no.~5, p. 054613, May 2019.

\bibitem{usangCorrelated2019}
M.~D. Usang, F.~A. Ivanyuk, C.~Ishizuka, and S.~Chiba, ``Correlated transitions
  in {{TKE}} and mass distributions of fission fragments described by 4-{{D
  Langevin}} equation,'' \emph{Sci Rep}, vol.~9, no.~1, pp. 1--9, Feb. 2019.

\bibitem{miyamotoOrigin2019}
Y.~Miyamoto, Y.~Aritomo, S.~Tanaka, K.~Hirose, and K.~Nishio, ``Origin of the
  dramatic change of fission mode in fermium isotopes investigated using
  {{Langevin}} equations,'' \emph{Phys. Rev. C}, vol.~99, no.~5, p. 051601, May
  2019.

\bibitem{parisiPerturbation1981}
G.~Parisi, Y.~S. Wu \emph{et~al.}, ``Perturbation theory without gauge
  fixing,'' \emph{Sci. Sin}, vol.~24, no.~4, pp. 483--496, 1981.

\bibitem{damgaardStochastic1984}
P.~Damgaard and K.~Tsokos, ``Stochastic quantization with fermions,''
  \emph{Nuclear Physics B}, vol. 235, no.~1, pp. 75--92, 1984.

\bibitem{sadhukhanMicroscopic2016}
J.~Sadhukhan, W.~Nazarewicz, and N.~Schunck, ``Microscopic modeling of mass and
  charge distributions in the spontaneous fission of $^{240}$pu,'' \emph{Phys.
  Rev. C}, vol.~93, no.~1, p. 011304, Jan. 2016.

\bibitem{bulgacFission2019}
A.~Bulgac, S.~Jin, K.~J. Roche, N.~Schunck, and I.~Stetcu, ``Fission dynamics
  of 240pu from saddle to scission and beyond,'' \emph{Phys. Rev. C}, vol. 100,
  no.~3, p. 034615, Sep. 2019.

\bibitem{mutherSingle1977}
\BIBentryALTinterwordspacing
H.~M\"uther, K.~Goeke, K.~Allaart, and A.~Faessler, ``Single-particle degrees
  of freedom and the generator-coordinate method,'' \emph{Phys. Rev. C},
  vol.~15, pp. 1467--1476, Apr 1977. [Online]. Available:
  \url{https://link.aps.org/doi/10.1103/PhysRevC.15.1467}
\BIBentrySTDinterwordspacing

\bibitem{chenTriaxial2017}
F.-Q. Chen and J.~L. Egido, ``Triaxial shape fluctuations and quasiparticle
  excitations in heavy nuclei,'' \emph{Phys. Rev. C}, vol.~95, no.~2, p.
  024307, Feb. 2017.

\bibitem{bulgacInduced2016}
A.~Bulgac, P.~Magierski, K.~J. Roche, and I.~Stetcu, ``Induced {{Fission}} of
  {\textsuperscript{240}}{{Pu}} within a {{Real}}-{{Time Microscopic
  Framework}},'' \emph{Phys. Rev. Lett.}, vol. 116, no.~12, p. 122504, Mar.
  2016.

\bibitem{bulgacUnitary2019}
\BIBentryALTinterwordspacing
A.~Bulgac, S.~Jin, and I.~Stetcu, ``Unitary evolution with fluctuations and
  dissipation,'' \emph{Phys. Rev. C}, vol. 100, p. 014615, Jul 2019. [Online].
  Available: \url{https://link.aps.org/doi/10.1103/PhysRevC.100.014615}
\BIBentrySTDinterwordspacing

\bibitem{sierkLangevin2017}
A.~J. Sierk, ``Langevin model of low-energy fission,'' \emph{Phys. Rev. C},
  vol.~96, no.~3, 2017.

\bibitem{wadaMulti1992}
\BIBentryALTinterwordspacing
T.~Wada, N.~Carjan, and Y.~Abe, ``Multi-dimensional langevin approach to
  fission dynamics,'' \emph{Nuclear Physics A}, vol. 538, pp. 283 -- 289, 1992.
  [Online]. Available:
  \url{http://www.sciencedirect.com/science/article/pii/037594749290778I}
\BIBentrySTDinterwordspacing

\bibitem{bernardTaking2011}
R.~Bernard, ``Taking into account the intrinsic excitations and their couplings
  to collective modes in the fission process; couplages modes collectifs -
  excitations intrinseques dans le processus de fission,'' Ph.D. dissertation,
  Université Pierre et Marie Curie, Sep 2011.

\bibitem{bernardMicroscopic2011}
R.~Bernard, H.~Goutte, D.~Gogny, and W.~Younes, ``Microscopic and nonadiabatic
  {{Schr{\"o}dinger}} equation derived from the generator coordinate method
  based on zero- and two-quasiparticle states,'' \emph{Phys. Rev. C}, vol.~84,
  no.~4, p. 044308, Oct. 2011.

\bibitem{younesMicroscopic2019}
\BIBentryALTinterwordspacing
W.~Younes, D.~M. Gogny, and J.-F. Berger, \emph{A {{Microscopic Theory}} of
  {{Fission Dynamics Based}} on the {{Generator Coordinate Method}}}, ser.
  Lecture {{Notes}} in {{Physics}}.\hskip 1em plus 0.5em minus 0.4em\relax
  {Springer International Publishing}, 2019. [Online]. Available:
  \url{//www.springer.com/us/book/9783030044220}
\BIBentrySTDinterwordspacing

\bibitem{holzwarthConnection1972}
G.~Holzwarth, ``The connection between the generator coordinate method and
  {{Bose}} expansions,'' \emph{Nuclear Physics A}, vol. 185, no.~1, pp.
  268--272, Apr. 1972.

\bibitem{kermanQuantum1974}
A.~K. Kerman and S.~E. Koonin, ``Quantum {{Theory}} of {{Dissipation}} for
  {{Nuclear Collective Motion}},'' \emph{Phys. Scr.}, vol.~10, no.~A, pp.
  118--121, Jan. 1974.

\bibitem{scampsSystematics2013}
\BIBentryALTinterwordspacing
G.~Scamps and D.~Lacroix, ``Systematics of isovector and isoscalar giant
  quadrupole resonances in normal and superfluid spherical nuclei,''
  \emph{Phys. Rev. C}, vol.~88, p. 044310, Oct 2013. [Online]. Available:
  \url{https://link.aps.org/doi/10.1103/PhysRevC.88.044310}
\BIBentrySTDinterwordspacing

\bibitem{scampsSystematic2014}
------, ``Systematic study of isovector and isoscalar giant quadrupole
  resonances in normal and superfluid deformed nuclei,'' \emph{Phys. Rev. C},
  vol.~89, no.~3, p. 034314, Mar. 2014.

\bibitem{meyerMulticonfigurational1990}
H.~D. Meyer, U.~Manthe, and L.~S. Cederbaum, ``The multi-configurational
  time-dependent {{Hartree}} approach,'' \emph{Chemical Physics Letters}, vol.
  165, no.~1, pp. 73--78, Jan. 1990.

\bibitem{mantheWave1992}
U.~Manthe, H.-D. Meyer, and L.~S. Cederbaum, ``Wave-packet dynamics within the
  multiconfiguration {{Hartree}} framework: {{General}} aspects and application
  to {{NOCl}},'' \emph{J. Chem. Phys.}, vol.~97, no.~5, pp. 3199--3213, Sep.
  1992.

\bibitem{beckMulticonfiguration2000}
M.~H. Beck, A.~J{\"a}ckle, G.~A. Worth, and H.~D. Meyer, ``The
  multiconfiguration time-dependent {{Hartree}} ({{MCTDH}}) method: A highly
  efficient algorithm for propagating wavepackets,'' \emph{Physics Reports},
  vol. 324, no.~1, pp. 1--105, Jan. 2000.

\bibitem{meyerMultidimensional2009}
H.-D. Meyer, F.~Gatti, and G.~A. Worth, \emph{Multidimensional {{Quantum
  Dynamics}}: {{MCTDH Theory}} and {{Applications}}}.\hskip 1em plus 0.5em
  minus 0.4em\relax {John Wiley \& Sons}, Apr. 2009.

\bibitem{ndengueStatetostate2019}
S.~Ndengu{\'e}, Y.~Scribano, F.~Gatti, and R.~Dawes, ``State-to-state inelastic
  rotational cross sections in five-atom systems with the multiconfiguration
  time dependent {{Hartree}} method,'' \emph{J. Chem. Phys.}, vol. 151, no.~13,
  p. 134301, Oct. 2019.

\bibitem{wangMultilayer2003}
H.~Wang and M.~Thoss, ``Multilayer formulation of the multiconfiguration
  time-dependent {{Hartree}} theory,'' \emph{J. Chem. Phys.}, vol. 119, no.~3,
  pp. 1289--1299, Jul. 2003.

\bibitem{nestMulticonfiguration2005}
M.~Nest, T.~Klamroth, and P.~Saalfrank, ``The multiconfiguration time-dependent
  {{Hartree}}\textendash{{Fock}} method for quantum chemical calculations,''
  \emph{J. Chem. Phys.}, vol. 122, no.~12, p. 124102, Mar. 2005.

\bibitem{hochstuhlTwophoton2011}
D.~Hochstuhl and M.~Bonitz, ``Two-photon ionization of helium studied with the
  multiconfigurational time-dependent {{Hartree}}\textendash{{Fock}} method,''
  \emph{J. Chem. Phys.}, vol. 134, no.~8, p. 084106, Feb. 2011.

\bibitem{katoTime2004}
T.~Kato and H.~Kono, ``Time-dependent multiconfiguration theory for electronic
  dynamics of molecules in an intense laser field,'' \emph{Chemical physics
  letters}, vol. 392, no. 4-6, pp. 533--540, 2004.

\bibitem{katoTime2008}
------, ``Time-dependent multiconfiguration theory for electronic dynamics of
  molecules in intense laser fields: {{A}} description in terms of numerical
  orbital functions,'' \emph{J. Chem. Phys.}, vol. 128, no.~18, p. 184102, May
  2008.

\bibitem{robinDescription2016}
C.~Robin, N.~Pillet, D.~Pe{\~n}a~Arteaga, and J.-F. Berger, ``Description of
  nuclear systems with a self-consistent configuration-mixing approach:
  {{Theory}}, algorithm, and application to the {{C}} 12 test nucleus,''
  \emph{Phys. Rev. C}, vol.~93, no.~2, Feb. 2016.

\bibitem{robinDescription2017}
C.~Robin, N.~Pillet, M.~Dupuis, J.~Le~Bloas, D.~Pe{\~n}a~Arteaga, and J.-F.
  Berger, ``Description of nuclear systems with a self-consistent
  configuration-mixing approach. {{II}}. {{Application}} to structure and
  reactions in even-even s d -shell nuclei,'' \emph{Phys. Rev. C}, vol.~95,
  no.~4, Apr. 2017.

\bibitem{orestesGenerator2007}
E.~Orestes, K.~Capelle, A.~B.~F. {da Silva}, and C.~A. Ullrich, ``Generator
  coordinate method in time-dependent density-functional theory: {{Memory}}
  made simple,'' \emph{J. Chem. Phys.}, vol. 127, no.~12, p. 124101, Sep. 2007.

\bibitem{reinhardTime1983}
P.~G. Reinhard, R.~Y. Cusson, and K.~Goeke, ``Time evolution of coherent
  ground-state correlations and the {{TDHF}} approach,'' \emph{Nuclear Physics
  A}, vol. 398, no.~1, pp. 141--188, Apr. 1983.

\bibitem{regnierMicroscopic2019}
D.~Regnier and D.~Lacroix, ``Microscopic description of pair transfer between
  two superfluid {{Fermi}} systems. {{II}}. {{Quantum}} mixing of
  time-dependent {{Hartree}}-{{Fock}}-{{Bogolyubov}} trajectories,''
  \emph{Phys. Rev. C}, vol.~99, no.~6, p. 064615, Jun. 2019.

\bibitem{scampsPairing2012}
G.~Scamps, ``Pairing dynamics in particle transport,'' \emph{Phys. Rev. C},
  vol.~85, no.~3, 2012.

\bibitem{scampsSuperfluid2017}
G.~Scamps and Y.~Hashimoto, ``Superfluid effects in collision between systems
  with small particle number,'' \emph{EPJ Web Conf.}, vol. 163, p. 00049, 2017.

\bibitem{scampsTransfer2017}
G.~Scamps, ``Transfer probabilities for the reactions,'' \emph{Phys. Rev. C},
  vol.~96, no.~3, 2017.

\bibitem{regnierMicroscopic2018}
D.~Regnier, D.~Lacroix, G.~Scamps, and Y.~Hashimoto, ``Microscopic description
  of pair transfer between two superfluid {{Fermi}} systems: {{Combining}}
  phase-space averaging and combinatorial techniques,'' \emph{Phys. Rev. C},
  vol.~97, no.~3, p. 034627, Mar. 2018.

\end{thebibliography}

\end{document}